\documentclass[11pt,prd,onecolumn,amsmath,amssymb,aps,floats,floatfix, nofootinbib]{revtex4-2}

\usepackage[colorlinks=true,urlcolor=blue,anchorcolor=blue,citecolor=blue,filecolor=blue,linkcolor=blue,menucolor=blue,linktocpage=true]{hyperref} 

\usepackage[inline]{enumitem}
\usepackage[multidot]{grffile}  
\usepackage{dcolumn}
\usepackage{bm}
\usepackage{amsmath}
\usepackage{amsfonts}
\usepackage{amssymb}
\usepackage{color}
\usepackage[table]{xcolor}
\usepackage{float}
\usepackage{latexsym}
\usepackage{slashed} 
\usepackage{pstricks}
\usepackage{indentfirst}
\usepackage{mathrsfs}
\usepackage{multirow}
\usepackage{epsfig,psfrag}
\usepackage{graphicx}
\usepackage{enumitem}
\usepackage{geometry,amssymb,yfonts}
\usepackage{yhmath}
\usepackage{anysize}
\usepackage{subfigure}
\usepackage{mathtools}
\usepackage{setspace} 
\usepackage[utf8]{inputenc} 
\usepackage[scientific-notation=true]{siunitx} 
\usepackage{feynmp-auto}
\usepackage{verbatim}

\usepackage[normalem]{ulem}
\usepackage{longtable}

\graphicspath{{fig/}}

\allowdisplaybreaks 

\begin{document}
	
\title{Probing Long-Lived Photophobic Axion-Like Particles via Prompt Leptons and Mono-$\gamma$ at FCC-ee and CEPC}

\author{Shi-Yu Wang$^{a}$}
\author{Yu-Peng Jiao$^{a}$}
\author{Hong-Hao Zhang$^{a}$}\email[Corresponding author. ]{zhh98@mail.sysu.edu.cn}
\author{Giacomo Cacciapaglia$^{b}$}\email[Corresponding author. ]{cacciapa@lpthe.jussieu.fr}

\affiliation{$^a$School of Physics, Sun Yat-Sen University, Guangzhou 510275, China}
\affiliation{$^b$Laboratoire de Physique Th\'eorique et Hautes \'Energies (LPTHE), UMR 7589,
Sorbonne Universit\'e \& CNRS, 4 place Jussieu, 75252 Paris Cedex 05, France}

\begin{abstract}
	We investigate the potential to probe axion-like particles (ALPs) under the photophobic scenario at the FCC-ee and CEPC at the Z-pole, with $\sqrt{s} = 91.2$ GeV. The signal process is
	$$
	e^+e^- \to Z \to \gamma a,\quad a \to \ell^+ \ell^-,
	$$
	where we consider final states with two prompt leptons and one photon, or only one photon (Mono-$\gamma$). We estimate the sensitivity to the ALP mass $m_a$ and associated energy scale $\Lambda$ for $a\to\mu^+\mu^-$ and $a\to\tau^+\tau^-$ (with leptonic decays of the $\tau$) by use of a XGBoost classifier. For an integrated luminosity of 150 ab$^{-1}$ at the Z-pole, the combined leptonic channel can probe the ALP scale $\Lambda$ between $10$ to $700$~TeV, depending on the ALP mass. The Mono-$\gamma$ signal offers a complementary probe, reaching $\Lambda$ up to $2000$~TeV for masses below $20$~GeV.
	
\end{abstract}

\maketitle
\tableofcontents

\section{Introduction}

Axions and axion-like particles (ALPs) are light gauge-singlet pseudoscalars that typically emerge as Goldstone bosons of a spontaneously broken global $U(1)$ symmetry at a high scale. They are predicted in many classes of models, leading to effective theories at the electroweak scale featuring various coupling structures: the common denominator being the presence of a topological coupling to gauge bosons and derivative couplings to fermions. Historically, the first instance where axions appeared is a solution to the strong CP problem in the Standard Model (SM).

In Quantum Chromodynamics (QCD), the Lagrangian exhibits a global symmetry $U(N_f)_V \times U(N_f)_A$ in the chiral limit where the quark masses vanish ($m_q \to 0$). Since $m_u, m_d \ll \Lambda_{\text{QCD}}$, it is expected that the strong interaction approximately respects a $U(2)_V \times U(2)_A$ symmetry. Experimentally, while the vector symmetry $U(2)_V = SU(2)_I \times U(1)_B$ (i.e. isospin and baryon number) is indeed a good approximate symmetry, the axial symmetry is found to be badly violated \cite{Peccei:2006as}. This discrepancy is known as the $U(1)_A$ problem, as first identified by Weinberg \cite{Weinberg:1975ui}.
The problem was resolved by t'Hooft \cite{tHooft:1976rip,tHooft:1976snw}, who demonstrated that the QCD vacuum possesses a nontrivial topological structure, leading to an additional term in the QCD Lagrangian: $\theta \frac{g^2}{32\pi^2} G^{\mu\nu}_a \tilde{G}_{a\mu\nu}.$
This term explicitly violates CP symmetry, and current experimental bounds on the neutron electric dipole moment place a stringent constraint on the parameter $\theta \lesssim 10^{-11}$ \cite{Baker:2006ts,Baluni:1978rf,Crewther:1979pi,Kaplan:2025bgy,Kim:2008hd}. More properly, the bound applies to a physical combination of $\theta$ and the phase of the quark masses, hence the required fine-tuning problem is referred to as the \emph{strong CP problem} \cite{Weinberg:1976hu}.
The first attempt to address the strong CP problem postulated a global $U(1)_{PQ}$ symmetry \cite{Peccei:1977hh,Peccei:1977ur}, hence canceling the static CP-violating angle ${\theta}$ with a dynamical, CP-conserving field—the axion \cite{Weinberg:1977ma,Wilczek:1977pj}. The axion emerges as the Nambu–Goldstone boson associated with the spontaneous breaking of the $U(1)_{PQ}$ symmetry. Under a $U(1)_{PQ}$ transformation, the axion field $a(x)$ shifts as $a(x) \to a(x) + \alpha f_a$. To preserve the $U(1)_{PQ}$ invariance of the SM Lagrangian, an axion coupling term must be introduced:
$\xi \frac{a}{f_a} \frac{g^2}{32\pi^2} G_a^{\mu\nu} \tilde{G}_{a\mu\nu}.$
This term also generates an effective potential for the axion, whose minimum lies at $\langle a \rangle = -\frac{f_a}{\xi} \theta$. The cancellation of the $\theta$ term at this minimum provides a dynamical solution to the strong CP problem \cite{Peccei:1977hh,Peccei:1977ur}. It also generates a mass for the axion proportional to $m_a \propto \Lambda_{\rm QCD}^2/f_a$, which is very small as the scale $f_a$ is required to be very large.

As discussed above, the axion is a theoretically well-motivated particle, but it may be only one representative of a broader family of new particles, ALPs. Such particles interact feebly with the SM fields, suppressed by the $U(1)$ symmetry breaking scale, but may receive sizable masses from interactions beyond the SM (typically masses ranging from sub-eV to the electroweak scale are expected). ALPs have attracted interests beyond their potential role in solving the CP problem, as they can be searched in various classes of experiments, with couplings to photons playing a key role for the lightest masses (see e.g. \cite{Graham:2015ouw,Chang:2000ii,Boehm:2014hva,Berlin:2014tja,Krasznahorkay:2015iga,Marciano:2016yhf,Feng:2016ysn,Ellwanger:2016wfe,Han:2022kvj,Mimasu:2014nea,Jaeckel:2015jla,Dolan:2017osp,Kleban:2005rj,Jiang:2018jqp,Belle-II:2020jti,Bauer:2021mvw,dEnterria:2021ljz,Ghebretinsaea:2022djg}). Meson properties can also be used, especially in the presence of flavor violating couplings  \cite{Izaguirre:2016dfi,Bauer:2021mvw,Alda:2025uwo}.

For larger masses, between the GeV and TeV scales, collider experiments have discovery potential, with the coupling to photon stealing the stage in most investigations \cite{Bauer:2017ris,Yue:2021iiu,Zhang:2021sio,Cacciapaglia:2021agf,Wang:2021uyb,Ren:2021prq,Polesello:2025gwj}. The photophobic ALP \cite{Craig:2018kne,Cai:2020bhd,Cai:2019cow} phenomenology at the LHC has also been studied \cite{Aiko:2024xiv,Ding:2024djo}. Future colliders, especially proposals involving lepton collisions, such as FCC-ee \cite{FCC:2018evy,FCC:2025lpp}, CEPC \cite{CEPCStudyGroup:2023quu} and a muon collider \cite{InternationalMuonCollider:2025sys}, also offer bright prospect on the discovery of ALPs \cite{Bauer:2018uxu}. 
In particular, Refs~\cite{Cacciapaglia:2021agf,Cacciapaglia:2022tfd} analyzed the sensitivity of a Tera-Z phase at a future electron-positron collider to ALPs under both the photophobic and photophilic scenarios, accounting for the effects of the ALP decay length (see also \cite{Polesello:2025gwj} for a more detailed collider simulation in the photophilic case). Finally, Refs~\cite{Bao:2025tqs,Bauer:2018uxu} explored the prospects of future lepton colliders to probe long-lived ALPs through mono-$\gamma$ and mono-$Z$ signatures.

This paper studies the discovery potential of ALPs in the photophobic scenario via prompt leptonic final states at the Z-pole at future colliders. We will consider prompt a decay length $L < 1$~cm, and ALP  masses below the $Z$ mass, $m_Z = 91.2$~GeV. In this regime, the ALP decay modes vary significantly with the mass: different fermion pairs dominate at different scales, while the diphoton channel is dominant for $m_a < 2m_e$ and $m_a \gtrsim 75$ GeV. At an electron-positron collider, such as FCC-ee and CEPC, with center-of-mass energy $\sqrt{s} = 91.2$ GeV, the process $Z \to \gamma a$, with the ALP decaying into a lepton pair, provides a particularly sensitive probe in the photophobic scenario. Since ALPs in this regime tend to be long-lived, many may escape the detector before decaying. We therefore compute their velocity and lifetime to estimate the decay length and determine the fraction of ALPs that decay promptly within the detector volume, which is critical for assessing the experimental sensitivity. In addition, by computing the lifetime, we find that axion-like particles with masses below $10$ GeV predominantly decay outside the detector, hence revealing a mono-photon signature. Therefore, we perform additional collider simulations specifically for this scenario. To overcome the model dependence in the couplings to fermions, we work under a specific scenario inspired by composite Higgs models, where only couplings to the electroweak gauge bosons are generated at the high scale $\Lambda$, while couplings to fermions (and photons) are generated at one loop level. All the couplings, therefore, are proportional to a single scale $1/\Lambda$.

Our analysis shows that, when the dominant ALP decay mode is into electrons, the prompt ALP signal is heavily suppressed. Therefore, we restrict our study to ALP decays into $\mu$ and $\tau$, with $\tau$ further decaying into $e$, $\mu$, and corresponding neutrinos. We employ an XGBoost model to enhance signal-over-background discrimination, using the energy-momentum and spatial distributions of visible final-state particles, as well as the missing momentum from invisible particles, as input features. We find that XGBoost significantly improves the separation between signal and background. From our analysis of the leptonic decay channels, the ALP coupling scale $\Lambda$ can be excluded up to $700$~TeV combining all the channels. For the Mono-$\gamma$ signal, the coupling can be excluded up to $2000$~TeV for masses below $20$~GeV.

The structure of this paper is as follows: Sec.~\ref{ALPmodel} introduces the theoretical framework of ALPs, including the relevant Lagrangian and coupling structures; Sec.~\ref{ALPdecay} discusses ALP production and decay behavior at the Z pole in the photophobic composite scenario, with emphasis on the impact of the ALP decay length on observability; Sec.~\ref{Simulation} details the simulation pipeline and the methodology for signal-over-background discrimination; Sec.~\ref{result} presents the sensitivity projections for ALP mass and coupling scale; finally, Sec.~\ref{conclusion} summarizes the main results and provides concluding remarks.

\section{ALP model}
\label{ALPmodel}

In this study, the effective Lagrangian for the ALP $a$, written before electroweak symmetry breaking (EWSB), is given by \cite{Bauer:2017ris}:
\begin{eqnarray}
	\begin{aligned}\mathcal{L}_{\mathrm{eff}}^{D\leq5}&\begin{aligned}&=\frac{1}{2}\left(\partial_{\mu}a\right)(\partial^{\mu}a)-\frac{m_{a,0}^{2}}{2}a^{2}+\frac{\partial^{\mu}a}{\Lambda}\sum_{F}\bar{\psi}_{F}C_{F}\gamma_{\mu}\psi_{F}\end{aligned}\\&+g_{s}^{2}C_{GG}\frac{a}{\Lambda}G_{\mu\nu}^{A}\tilde{G}^{\mu\nu,A}+g^{2}C_{WW}\frac{a}{\Lambda}W_{\mu\nu}^{a}\tilde{W}^{\mu\nu,a}+g^{\prime2}C_{BB}\frac{a}{\Lambda}B_{\mu\nu}\tilde{B}^{\mu\nu}\,,\end{aligned}
	\label{L1}
\end{eqnarray}
where $C_F$ denotes the ALP–fermion coupling coefficient, while $C_{XX}$ indicate the coupling to the SM gauge bosons, where $G_{\mu\nu}^A$, $W_{\mu\nu}^a$, and $B_{\mu\nu}$ are the field strength tensors of the $SU(3)_c$, $SU(2)_L$, and $U(1)_Y$ gauge groups, respectively, with coupling constants $g_s$, $g$, and $g'$. The dual field strength is defined as $\tilde{B}^{\mu\nu} = \frac{1}{2} \epsilon^{\mu\nu\alpha\beta} B_{\alpha\beta}$, with $\epsilon^{0123} = 1$.

After EWSB, the effective Lagrangian~\ref{L1} generates couplings of the ALP to mass eigenstates $\gamma\gamma$, $\gamma Z$, and $ZZ$, which are given by:
\begin{eqnarray}
	\mathcal{L}_{\mathrm{eff}}^{D\leq5}\supset e^2C_{\gamma\gamma}\:\frac{a}{\Lambda}\:F_{\mu\nu}\:\tilde{F}^{\mu\nu}+\frac{2e^2}{s_wc_w}\:C_{\gamma Z}\:\frac{a}{\Lambda}\:F_{\mu\nu}\:\tilde{Z}^{\mu\nu}+\frac{e^2}{s_w^2c_w^2}\:C_{ZZ}\:\frac{a}{\Lambda}\:Z_{\mu\nu}\:\tilde{Z}^{\mu\nu}\:,
\end{eqnarray}
where $s_w \equiv \sin\theta_w$ and $c_w \equiv \cos\theta_w$, with coefficients defined as follows:
\begin{eqnarray}
	C_{\gamma\gamma}=C_{WW}+C_{BB}\:,\quad C_{\gamma Z}=c_w^2\:C_{WW}-s_w^2\:C_{BB}\quad C_{ZZ}=c_w^4\:C_{WW}+s_w^4\:C_{BB}\:.
\end{eqnarray}
Note that all the couplings are normalized by a common scale $\Lambda$.

Fermion mass terms are diagonalized after EWSB via field redefinitions such that, for instance, $U_u^\dagger Y_u W_u = \mathrm{diag}(y_u, y_c, y_t)$. Under this basis rotation, the fermionic coupling matrix $C_F$ is transformed as \cite{Bauer:2017ris}:
\begin{eqnarray}
	{K}_{U}={U}_{u}^{\dagger}{C}_{Q}{U}_{u}\mathrm{~,~}\quad{K}_{D}={U}_{d}^{\dagger}{C}_{Q}{U}_{d}\mathrm{~,~}\quad{K}_{E}={U}_{e}^{\dagger}{C}_{L}{U}_{e}\mathrm{~,~}\quad
	K_f=W_f^\dagger C_f W_f\,,
\end{eqnarray}
where $U,D,E$ labels the couplings of the left-handed fields and $f=u,d,e$ the right-handed ones.
The interaction terms in the Lagrangian can be rewritten in the mass eigenstate basis:
\begin{eqnarray}
	\mathcal{L}_{\mathrm{eff}}^{D\leq5}\supset\sum_f\frac{c_{ff}}2\frac{\partial^\mu a}\Lambda\:\bar{f}\gamma_\mu\gamma_5f\,,
\end{eqnarray}
where the summation runs over all fermion mass eigenstates. The coefficients $c_{ff}$ are defined as matrices in flavor space, with $i = 1, 2, 3$:
\begin{eqnarray}
	c_{u_iu_j}=(K_u)_{ij}-(K_U)_{ij}\:,\quad c_{d_id_j}=(K_d)_{ij}-(K_D)_{ij}\:,\quad c_{e_ie_j}=(K_e)_{ii}-(K_E)_{ij}\:.
\end{eqnarray}
Note that flavor off-diagonal couplings are also generated via the diagonalization and one-loop renormalization running \cite{Bauer:2021mvw}, generating very strong bounds for ALP masses below $10$~GeV, where they can appear in meson decays \cite{Izaguirre:2016dfi,Bauer:2021mvw,Alda:2025uwo}. For simplicity we will not consider them in detail here, and only focus on flavor diagonal ones.

\subsection{Composite photophobic ALP}

Composite Higgs models have been proposed since the 80's as an attractive solution to the Higgs naturalness problem, based on the color dynamics observed in QCD \cite{Weinberg:1975gm,Susskind:1978ms,Dimopoulos:1979es}. In the most popular incarnations, the Higgs emerges as a pseudo Nambu-Goldstone boson from the spontaneous breaking of a chiral symmetry, like pions in QCD \cite{Kaplan:1983fs,Kaplan:1983sm,Contino:2003ve}. Furthermore, the (heaviest) SM fermions acquire mass via partial compositeness \cite{Kaplan:1991dc}. For more details on such models, we refer the reader to the reviews \cite{Panico:2015jxa,Cacciapaglia:2020kgq} and references therein.

One feature of most composite Higgs models is the presence of additional pseudo Nambu-Goldstone bosons, stemming from the same symmetry breaking pattern that generates the Higgs and top partners \cite{Mrazek:2011iu,Ferretti:2013kya}. An interesting and predictive class of models is based on an underlying gauge-fermion dynamics \cite{Ferretti:2013kya,Vecchi:2015fma}, where two species of confining fermions carry electroweak and QCD charges, respectively. Hence, ALPs emerge in two ways: from the breaking of global $U(1)$ symmetries \cite{Ferretti:2016upr,Belyaev:2016ftv,Cacciapaglia:2019bqz,BuarqueFranzosi:2021kky}, and from extended symmetry in the electroweak sector \cite{Arbey:2015exa,Cacciapaglia:2021agf}. The latter is of interest, as the resulting light ALP lacks couplings to gluons (hence evading most searches at hadron colliders \cite{Arbey:2015exa}), and it allows for a photophobic scenario \cite{Cacciapaglia:2021agf}.

In our analysis, we focus on the photophobic scenario defined by $C_{WW} = 1$, $C_{BB} = -1$, $C_{GG} = 0$, and $c_{ff} = 0$. The model is thus fully characterized by the ALP mass $m_a$ and the coupling scale $\Lambda$. Despite the vanishing leading-order fermionic and gluonic couplings, the ALP still couples to photons and fermions via one-loop effects. It should be noted that the scale $\Lambda$ is related to the composite Higgs decay constant $f_H$ at one loop, $\Lambda \sim 16 \pi^2 f_H$, hence it is expected to be around $100$~TeV for the natural value $f_H \sim 1$~TeV. 

While at the compositeness scale only couplings to the electroweak gauge bosons are generated, the effective ALP–fermion coupling $c_{ff}^{\mathrm{eff}}$ receives radiative contributions and, at the scale $\mu$, it is given by \cite{Bauer:2017ris}:
\begin{eqnarray}
	\begin{aligned}c_{ff}^{\mathrm{eff}}&=c_{ff}(\mu)-12Q_{f}^{2}\alpha^{2}C_{\gamma\gamma}\left[\operatorname{ln}\frac{\mu^{2}}{m_{f}^{2}}+g(\tau_{f})\right]\\&-\frac{3\alpha^2}{s_w^4}C_{WW}\left(\ln\frac{\mu^2}{m_W^2}+\frac{1}{2}\right)-\frac{12\alpha^2}{s_w^2c_w^2}C_{\gamma Z}Q_f\left(T_3^f-2Q_f s_w^2\right)\left(\ln\frac{\mu^2}{m_Z^2}+\frac{3}{2}\right)\\&-\frac{12\alpha^2}{s_w^4c_w^4}C_{ZZ}\left(Q_f^2s_w^4-T_3^f Q_f s_w^2+\frac{1}{8}\right)\left(\ln\frac{\mu^2}{m_Z^2}+\frac{1}{2}\right).\end{aligned}
\end{eqnarray}
Here, $c_{ff}(\mu)$ denotes the scale-dependent tree-level coupling of the ALP to leptons, $T_3^f$ is the third component of the fermion's weak isospin, and $Q_f$ is the electric charge of the fermion. The full integral expression for the loop function $g(\tau_f)$ is given by:
\begin{eqnarray}
	g(\tau)=5+\frac{4}{3}\int_0^1dx\frac{1-4\tau(1-x)^2-2x+4x^2}{\sqrt{\tau(1-x)^2-x^2}}\arctan\left(\frac{x}{\sqrt{\tau(1-x)^2-x^2}}\right)\;,
\end{eqnarray}
where $\tau = 4m_f^2 / m_a^2$. However, in hadronic physics, due to QCD confinement, a light ALP can only decay into hadronic final states. Moreover, the allowed decay modes are highly constrained in the mass region below $1$~GeV by CP and angular momentum conservation. In this regime, the dominant decay channels are $3\pi^0$ and $\pi^+\pi^-\pi^0$. For sufficiently light ALPs, the partial width for the decay into three pions can be expressed as
\begin{eqnarray}
	\Gamma(a\to\pi^a\pi^b\pi^0)=\frac{\pi}{6}\frac{m_am_\pi^4}{\Lambda^2f_\pi^2}\left[C_{GG}\frac{m_d-m_u}{m_d+m_u}+\frac{c^{\rm eff}_{uu}-c^{\rm eff}_{dd}}{32\pi^2}\right]^2g_{ab}\left(\frac{m_\pi^2}{m_a^2}\right),
\end{eqnarray}
where $f_\pi$ denotes the pion decay constant, $a$ and $b$ label the electric charges of the pions, and the phase-space function $g_{ab}$ is given by
\begin{eqnarray}
	g_{00}(r)&=\frac{2}{(1-r)^2}\int_{4r}^{(1-\sqrt{r})^2}dz\sqrt{1-\frac{4r}{z}}\lambda^{1/2}(1,z,r),\\
	g_{+-}(r)&=\frac{12}{(1-r)^2}\int_{4r}^{(1-\sqrt{r})^2}dz\sqrt{1-\frac{4r}{z}}(z-r)^2\lambda^{1/2}(1,z,r),
\end{eqnarray}
with $r = m_\pi^2/m_a^2$. When the ALP mass exceeds the QCD scale $\Lambda_{\rm QCD}$, the inclusive decay width into hadrons can be approximated as
\begin{eqnarray}
	\Gamma(a\to{\rm had.})=\frac{32\pi\alpha_s^2(m_a)m_a^3}{\Lambda^2}\left[1+\left(\frac{97}{4}-\frac{7n_q}{6}\right)\frac{\alpha_s(m_a)}{\pi}\right]\left|C_{GG}+\sum_{q=1}^{3}\frac{c_{qq}}{32\pi^2}\right|^2.
\end{eqnarray}
In the mass range between 1 and 3~GeV, a large number of exclusive hadronic channels become kinematically accessible, making it difficult to compute the individual branching ratios reliably due to non-perturbative QCD effects. Therefore, we omit this intermediate mass region in our subsequent analysis and simulations.

The effective coupling of the ALP to photons, $C_{\gamma\gamma}^{\mathrm{eff}}$, can be written as:
\begin{eqnarray}
	C_{\gamma\gamma}^\mathrm{eff}(m_a\gg\Lambda_\mathrm{QCD})=C_{\gamma\gamma}+\sum_f\frac{N_c^fQ_f^2}{16\pi^2}c_{ff}B_1(\tau_f)+\frac{2\alpha}{\pi}\frac{C_{WW}}{s_w^2}B_2(\tau_W)\;,
\end{eqnarray}
where, $\tau_i \equiv 4m_i^2 / m_a^2$, and $N_c^f$ and $Q_f$ denote the color factor and electric charge (in units of $e$) of the fermion $f$, respectively. The loop function is defined as:
\begin{eqnarray}
	\left.\begin{aligned}&B_1(\tau)=1-\tau\:f^2(\tau)\:,\\&B_2(\tau)=1-\left(\tau-1\right)f^2(\tau)\:,\end{aligned}\quad\text{where}\quad f(\tau)=\left\{\begin{array}{ll}\arcsin\frac1{\sqrt{\tau}}\:;&\quad\tau\geq1\:,\\\frac\pi2+\frac i2\ln\frac{1+\sqrt{1-\tau}}{1-\sqrt{1-\tau}}\:;&\quad\tau<1\:.\end{array}\right.\right.
\end{eqnarray}
The partial decay width of the ALP into two photons is given by:
\begin{eqnarray}
\Gamma(a \to \gamma\gamma) \equiv \frac{4\pi\alpha^2 m_a^3}{\Lambda^2} \left|C_{\gamma\gamma}^\mathrm{eff}\right|^2\,
\end{eqnarray}
which scales with the cube of the ALP mass. 

\section{ALP production and decay}
\label{ALPdecay}

\begin{figure}
	\centering
	\includegraphics[width=0.5\linewidth]{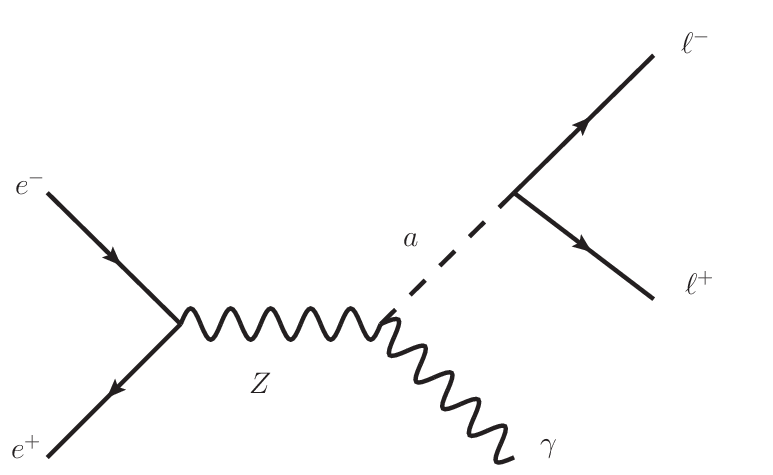}
	\caption{Feynman diagrams for $e^+e^- \to Z \to \gamma a,\quad a \to \ell^+ \ell^-$.}
	\label{fig:feiman}
\end{figure}

This study aims to evaluate the sensitivity of FCC-ee and CEPC to photophobic ALPs during the high-luminosity Z-pole operation. A widely studied channel is the decay of the Z boson into a photon and an ALP, with the ALP subsequently decaying; the corresponding Feynman diagram is shown in Fig.\ref{fig:feiman}. In the following collider simulation, we require the ALP to be on-shell. Therefore, the cross section for the process can be computed using the narrow width approximation: $
\sigma(e^+e^- \to \gamma \ell^+ \ell^-) = \sigma(e^+e^- \to \gamma a) \times \text{BR}(a \to \ell^+ \ell^-)$.

\begin{figure}
	\centering
	\includegraphics[width=0.7\linewidth]{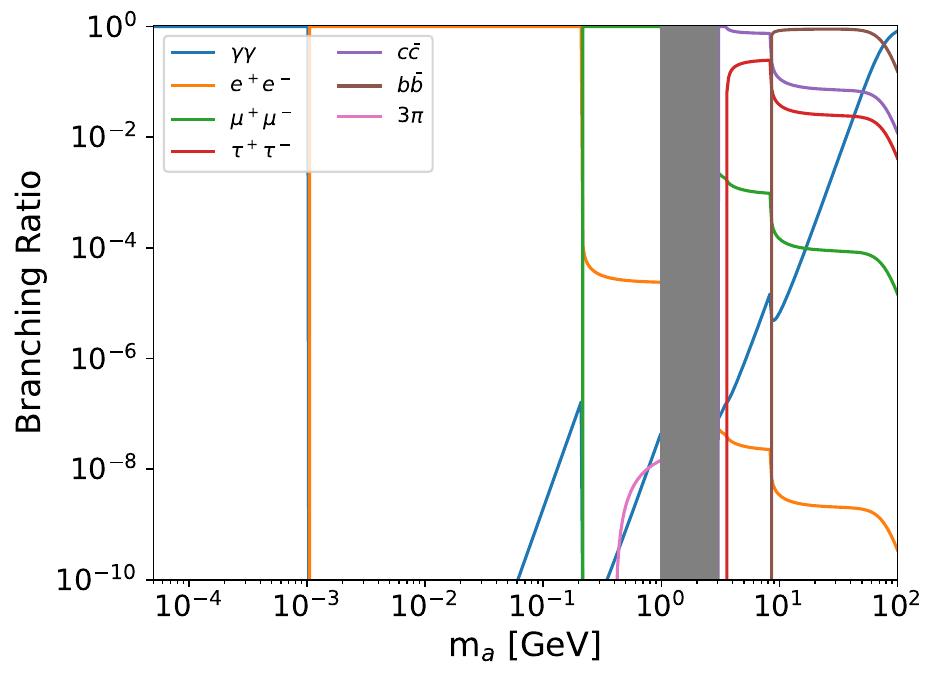}
	\caption{Branching ratios of the ALP decay channels as a function of the mass in the photophobic composite scenario ($C_{WW}=1$, $C_{BB}=-1$, $C_{GG}=0$, $c_{ff}=0$).}
	\label{fig:decay}
\end{figure}

The decay modes for this ALP are very sensitive to the mass $m_a$. To examine this behavior in detail, we compute the ALP decay branching fractions using the formulae provided in Ref.\cite{Bauer:2017ris}, and show the results in Fig.\ref{fig:decay}. In the mass range $[2m_\mu, 2m_c]$, the ALP primarily decays into a $\mu^+ \mu^-$ pair. For $m_a \in [2m_c, 2m_b]$, the dominant decay channels shift to $c\bar{c}$ jets and $\tau^+ \tau^-$ leptons. In the range $[2m_b, 75]$ GeV, the final state $b\bar{b}$ becomes the leading decay mode, with the $c\bar{c}$ and $\tau^+ \tau^-$ modes contributing subdominantly.
The photon channel has a very suppressed partial width, which also scales like $m_a^3$, hence it will only dominate for masses below $m_a < 2 m_e$ and at masses greater than $75$~GeV.

In this work, we focus on ALPs decaying via the $\mu^+ \mu^-$ and $\tau^+ \tau^-$ channels within the prompt region (defined as a decay length $L < 1$~cm), for ALP masses in the range $m_a \in [0.001, 91.2]$~GeV. Under the photophobic scenario considered, the ALP behaves as a long-lived particle (LLP), with a finite decay length inside the detector volume. The lab-frame decay length is given by:
\begin{eqnarray}
	L_{a} = \frac{E_{a}}{m_{a}\Gamma_{a}} \hbar c\,,
\end{eqnarray}
where $E_a = \frac{m_Z^2 + m_a^2}{2 m_Z}$ is the ALP energy from the $Z \to \gamma a$ decay. We take $\hbar c \approx 1.973 \times 10^{-13}$~GeV·cm, and $\Gamma_a$ denotes the total decay width of the ALP, which is computed by summing over all kinematically accessible final states. The probability that an ALP decays within the prompt region is then evaluated using the exponential decay law described previously:
\begin{eqnarray}
	P_{\text{decay}} (L_{\text{min}},L_{\text{max}}) = e^{-L_{\text{min}}/L_a} - e^{-L_{\text{max}}/L_a}\;.
\end{eqnarray}
Specifically, we define the prompt decay probability $P_{\text{prompt}}$ as the probability that the ALP decays within a radial distance between $L_{\text{min}} = 0$~cm and $L_{\text{max}} = 1$~cm from the interaction point.

\begin{figure}
	\centering
	\includegraphics[width=0.7\linewidth]{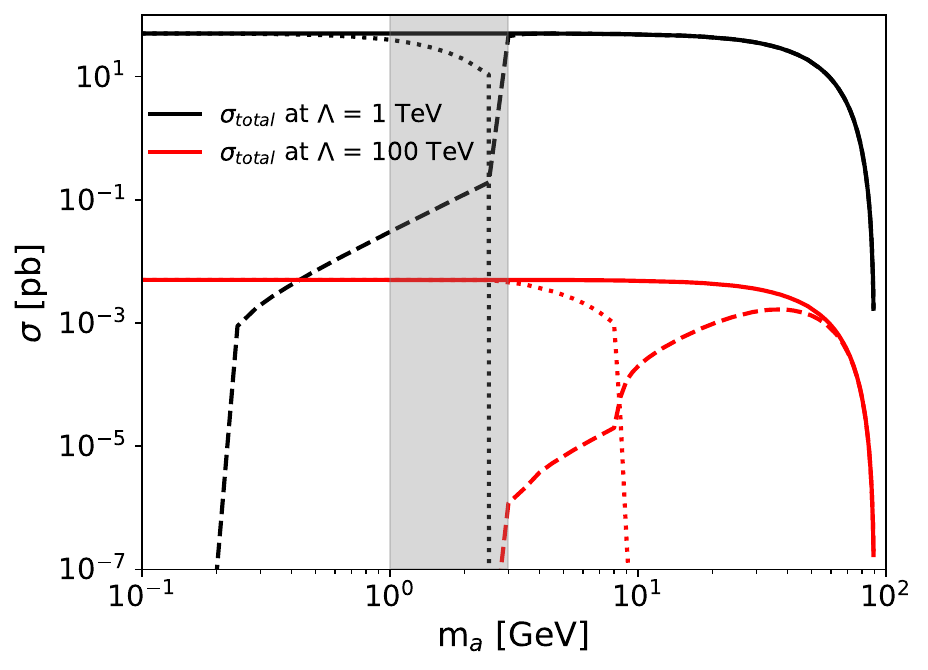}
	\caption{In the photophobic scenario, the variation of the cross section for the process $e^+e^- \to \gamma a$ is shown for different values of $\Lambda$. The solid lines represent the total cross section assuming the ALP decays anywhere. The dashed lines correspond to the cross section for prompt ALP decays ($P_{\text{decay}}(0,1~\text{cm})$), while the dotted lines indicate the cross section for ALPs decaying outside the detector ($P_{\text{decay}} (450~\text{cm},\infty)$). The black and red lines correspond to energy scales $\Lambda = 1$~TeV and $100$~TeV, respectively. The gray-shaded region between $1$ and $3$~GeV indicates the mass range in which exclusive hadronic decay channels (difficult to calculate) become kinematically accessible. In this region, we do not include the contribution of hadronic decays to the ALP lifetime.}
	\label{cs_prompt}
\end{figure}

To illustrate the impact of $m_a$ and $\Lambda$ on the prompt decay yield, we plot both the total cross section $\sigma(e^+ e^- \to \gamma a)$ (solid lines) and the prompt cross section $\sigma_{\text{prompt}}(e^+ e^- \to \gamma a) = \sigma(e^+ e^- \to \gamma a) \times P_{\text{prompt}}$ (dashed lines) as functions of the ALP mass in Fig.\ref{cs_prompt}. The energy scale $\Lambda$ is varied across $1$~TeV and $100$~TeV to reflect different coupling strengths.
As shown in Fig.\ref{cs_prompt}, for small $m_a$ and large $\Lambda$, the ALP lifetime becomes significantly longer, which suppresses the probability of decay within the prompt region. This leads to a substantial reduction of $\sigma_{\text{prompt}}$ compared to the total cross section.
Notably, $\sigma_{\text{prompt}}$ exhibits sharp increases near $m_a \approx 3$~GeV and $m_a \approx 9$~GeV. These features arise from the opening of the $\tau^+ \tau^-$ and $b\bar{b}$ decay channels, respectively, which drastically shorten the ALP lifetime and enhance the prompt decay probability.

In our simulation and analysis, we define the effective cross section for prompt ALP decays as 
\begin{eqnarray}
	\sigma_{\text{eff}} = P_{\text{prompt}} \times \sigma_{\text{total}} \,
\end{eqnarray} 
which reflects the production rate of ALPs decaying within the prompt region. As illustrated in Figs.\ref{fig:decay} and \ref{cs_prompt}, when the $a \to e^+e^-$ channel dominates, the effective cross section $\sigma_{\text{eff}}$ becomes heavily suppressed due to the long ALP lifetime in this regime. Compared to other decay channels, the prompt decay probability is significantly lower, rendering the signal effectively undetectable. For this reason, we do not consider the electron final states in our analysis. In Table~\ref{tab:data}, we present numerical values of the prompt decay probability $P_{\text{prompt}}$ for several benchmark ALP masses, assuming $\Lambda = 100$~TeV. For ALP masses below 10 GeV, we find that the Mono-$\gamma$ signal dominates. Therefore, we perform dedicated collider simulations for the signal processes in this mass range. The effective cross section for the Mono-$\gamma$ signal is shown in Fig.~\ref{cs_prompt} (dotted lines), where we consider the probability for the ALP to decay more than $450$~cm from the interaction point.

\section{Simulation detail}
\label{Simulation}

To evaluate the sensitivity to ALPs decaying into $\mu^+\mu^-$ and $\tau^+\tau^-$ final states, we employ the following Monte Carlo simulation chain. The hard-scattering events are generated using MadGraph5\_aMC@NLO \cite{Frederix:2018nkq}, followed by $\tau$ lepton decays simulated with PYTHIA 8.2 \cite{Bierlich:2022pfr}. Detector effects are modeled using DELPHES 3.5.0 \cite{deFavereau:2013fsa}, with the FCC-ee detector configuration card.

Since $\tau$ leptons decay promptly within the detector, we further assume they decay leptonically into $e$ or $\mu$, with branching ratios of 17.8\% and 17.4\%, respectively \cite{ParticleDataGroup:2024cfk}. As a result, the $\tau^+\tau^-$ channel leads to three distinct lepton-flavor combinations in the final state: $ee$, $\mu\mu$, and $e\mu$. Together with the direct $\mu^+\mu^-$ decay mode, we thus simulate four signal processes in total:

\begin{itemize}
\item $e^{+}e^{-} ~\to ~Z~ \to~ \gamma a ~,~ a ~\to~ \mu^{+} \mu^{-}$;
\item $e^{+}e^{-} ~\to ~Z~ \to~ \gamma a ~,~ a ~\to~ \tau^{+} \tau^{-} ~,~ \tau^{+} \to e^{+}\nu_{e}\bar{\nu}_{\tau} ~,~ \tau^{-} \to e^{-}\bar{\nu}_{e}\nu_{\tau}$;
\item $e^{+}e^{-} ~\to ~Z~ \to~ \gamma a ~,~ a ~\to~ \tau^{+} \tau^{-}~,~ \tau^{+} \to \mu^{+}\nu_{\mu}\bar{\nu}_{\tau} ~,~ \tau^{-} \to \mu^{-}\bar{\nu}_{\mu}\nu_{\tau}$;
\item $e^{+}e^{-} ~\to ~Z~ \to~ \gamma a ~,~ a ~\to~ \tau^{+} \tau^{-}~,~ \tau^{\pm} \to e^{\pm}\tilde{\nu}_{e}\tilde{\nu}_{\tau} ~,~ \tau^{\mp} \to \mu^{\mp}\tilde{\nu}_{\mu}\tilde{\nu}_{\tau}$;
\end{itemize}
where we indicate $\tilde{\nu} = \nu$ or $\bar \nu$.

Each of the four signal processes is associated with a corresponding irreducible SM background:

\begin{itemize}
	\item $e^{+}e^{-}  ~ \to~ \gamma \mu^{+} \mu^{-}$;
	\item $e^{+}e^{-} ~ \to~ \gamma \tau^{+} \tau^{-}$. 
\end{itemize}
We do not consider reducible backgrounds, stemming for instance from hadronic activity, as they are expected to be negligible and as they crucially depend on the detector characteristic.

Table~\ref{tab:data} presents the parton-level cross sections $\sigma_{\text{total}}$ for several ALP mass benchmarks, along with the branching ratios for $a\to\mu^+\mu^-$ and $a\to\tau^+\tau^-$, the prompt decay probability $P_{\text{prompt}}$, and the resulting effective cross section $\sigma_{\text{eff}}$ within the prompt region.

The benchmark scale $\Lambda = 100$~TeV is chosen to emphasize the impact of $P_{\text{prompt}}$ suppression at high $\Lambda$. The cross sections for the relevant SM backgrounds are $\sigma(e^+e^- \to \gamma\mu^+\mu^-) = 87$~pb and $\sigma(e^+e^- \to \gamma\tau^+\tau^-) = 84$~pb.

In evaluating these cross sections, we impose a set of parton-level selection cuts to reflect detector acceptance and trigger efficiency, as detailed in Refs.\cite{Bai:2025buy,Ripellino:2025dvg,FCC:2018evy}:

\begin{eqnarray}
	p_{T} (\ell,\gamma) &> 2 ~\text{GeV} \,, ~~ |\eta (\ell,\gamma)| < 2.5 \,.
\end{eqnarray}

\begin{table}[h!]
	\centering
	\begin{tabular}{c|cccccc}
		\toprule
		$m_{a}$~[GeV] & \textbf{10} & \textbf{30} & \textbf{50} & \textbf{70} & \textbf{80} & \textbf{88} \\
		\hline
		$\sigma_{\text{total}}^{e^{+}e^{-} \to \gamma a}$ [pb] & 0.0048 & 0.0035 & 0.0017 & 0.0003 & 6.11$\times$10$^{-5}$ & 1.2$\times$10$^{-6}$\\
		\hline
		$BR(a\to\mu^{+}\mu^{-})$ & 0.00014 & 8.4 $\times$10$^{-5}$ & 7.3$\times$10$^{-5}$ & 4.9$\times$ 10$^{-5}$ & 3.4$\times$10$^{-5}$  & 2.4$\times$10$^{-5}$\\
		\hline
		$BR(a\to\tau^{+}\tau^{-})$  &  0.038 & 0.024 & 0.020 & 0.014 & 0.0097 &0.0068 \\
		\hline
		$P_{\text{prompt}}$  &  0.041 & 0.45 & 0.80 &  0.98 & 1.0 & 1.0 \\
		\hline
		$\sigma_{\text{eff}}^{e^{+}e^{-} \to \gamma a}$ [pb] & 1.9$\times$10$^{-5}$ & 0.0015 &  0.0013 &  0.00029 & 6.1$\times$10$^{-5}$ & 1.16$\times$10$^{-6}$ \\
		\hline
	\end{tabular}
	\caption{Benchmark values of ALP total cross sections, branching ratios, prompt decay probabilities, and effective cross sections at $\Lambda = 100$~TeV.
	}
    	\label{tab:data}
\end{table}

\subsection{$\mu^{+}\mu^{-}$ channel}

\begin{figure}[t!]
	\centering
	\includegraphics[width=0.45\linewidth]{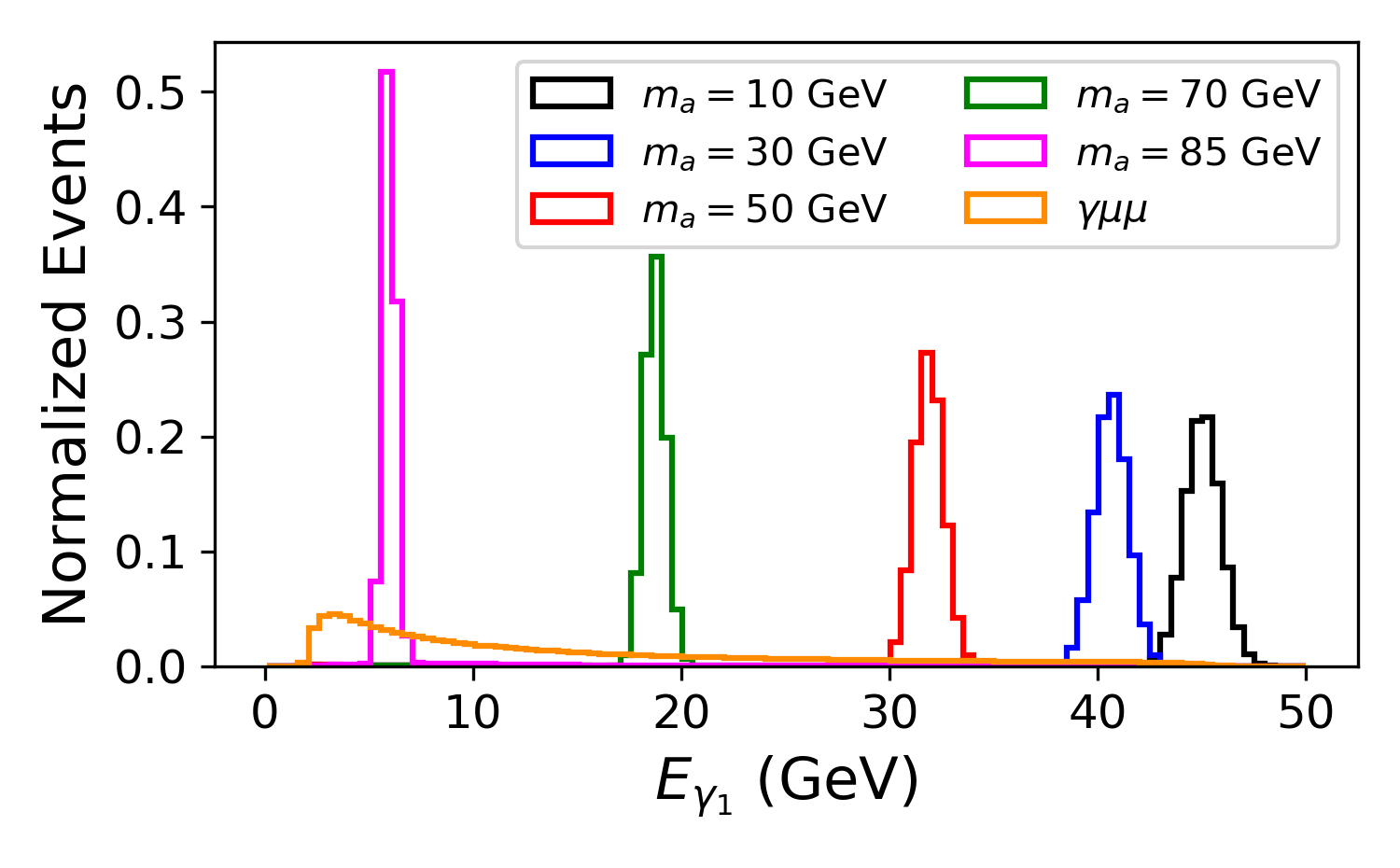}
	\includegraphics[width=0.45\linewidth]{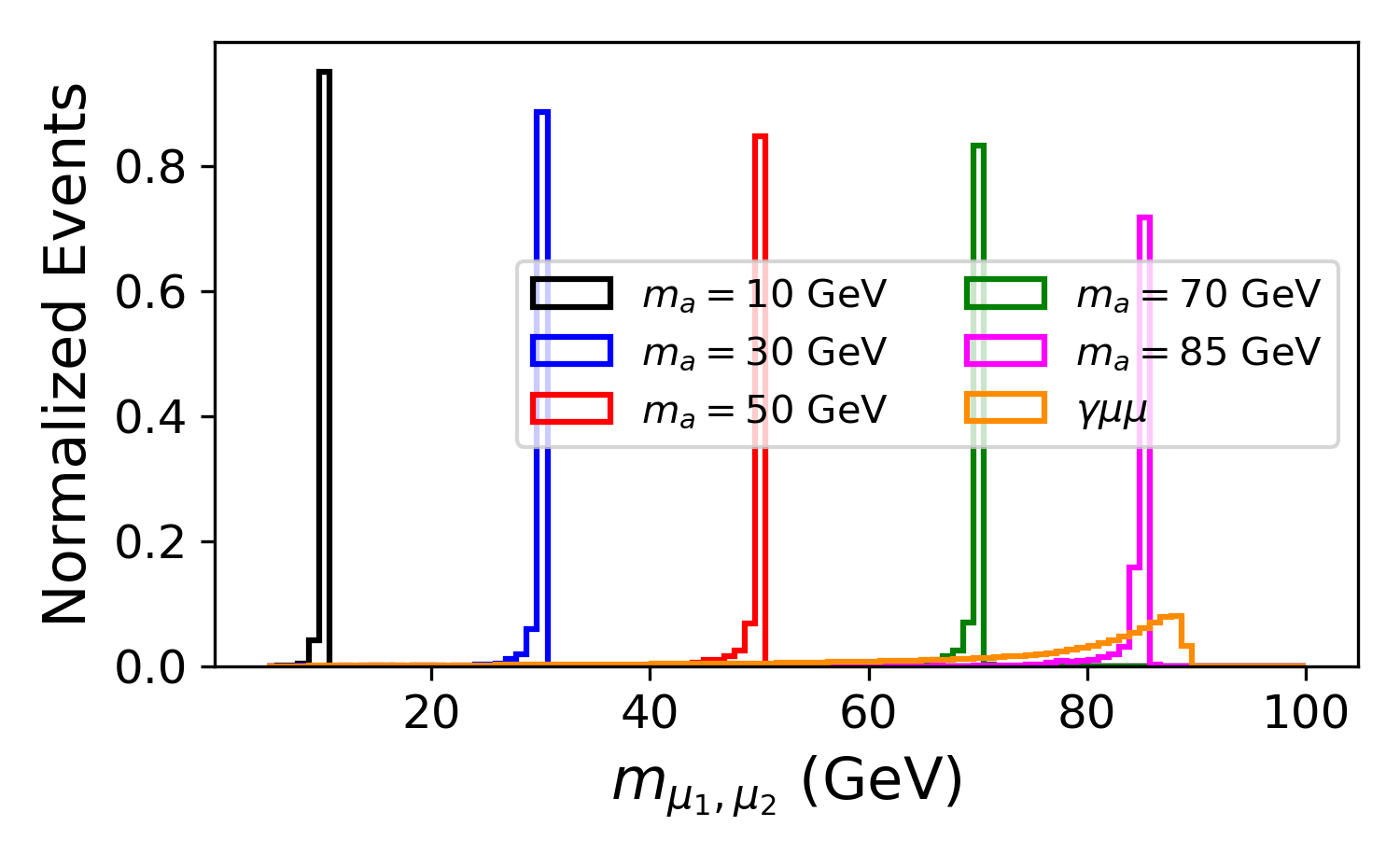}
	\caption{The left panel shows the normalized event distribution of the photon energy, while the right panel presents the reconstructed invariant mass of the ALP using the $\mu_1\mu_2$ system. The muons are ordered by decreasing $p_T$. Signal distributions for different ALP mass benchmarks are shown as solid lines, and the background is represented by dashed lines.}
	\label{fig:Egammaandma}
\end{figure}

At the detector level, the observables used in our analysis include the transverse momentum ($p_T$), pseudorapidity ($\eta$), and azimuthal angle ($\phi$) of the final-state muons and photon. In the signal process, the photon originates from the decay $Z \to \gamma a$, and its energy can be determined by the recoil relation:
\begin{eqnarray}
	E=\frac{E_{CM}^2-m_a^2}{2E_{CM}}\,,
	\label{Ea}
\end{eqnarray}
where $E_{\text{CM}} = m_Z = 91.2$~GeV is the center-of-mass energy at the $Z$-pole. The resulting photon energy spectrum, along with the reconstructed invariant mass of the muon pair, is shown in Fig.\ref{fig:Egammaandma}, where we order the muons by decreasing transverse momentum, $p_T$. The invariant mass $m_{\mu\mu}$, reconstructed from $\mu_1$ and $\mu_2$, serves as a key discriminating variable in our analysis. Since the ALP is treated as a long-lived particle with a narrow decay width, the signal distribution in $m_{\mu\mu}$ is sharply peaked near $m_a$. In contrast, the SM background exhibits a much broader distribution.

\begin{figure}[t!]
	\centering
	\includegraphics[width=0.45\linewidth]{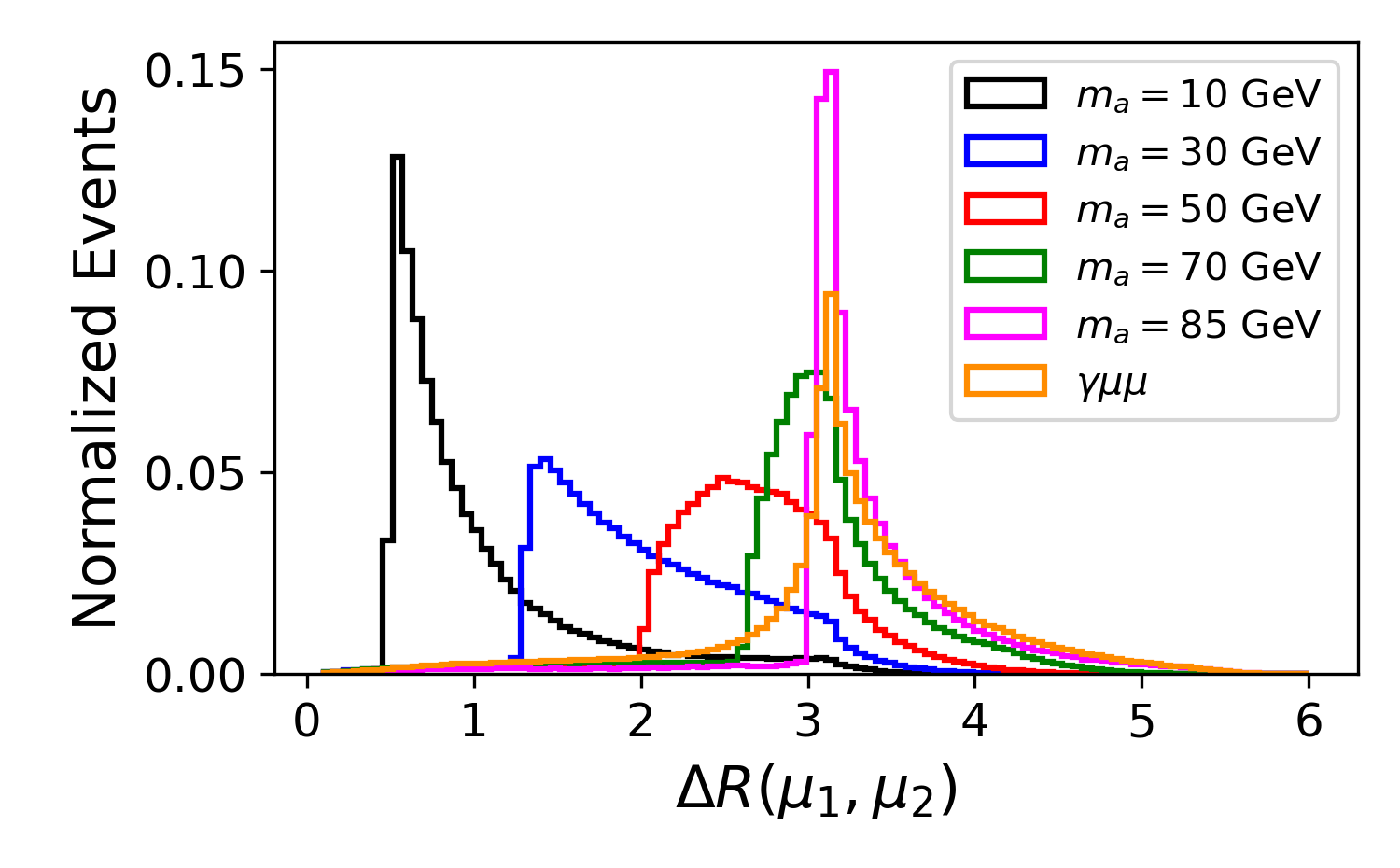}
	\includegraphics[width=0.45\linewidth]{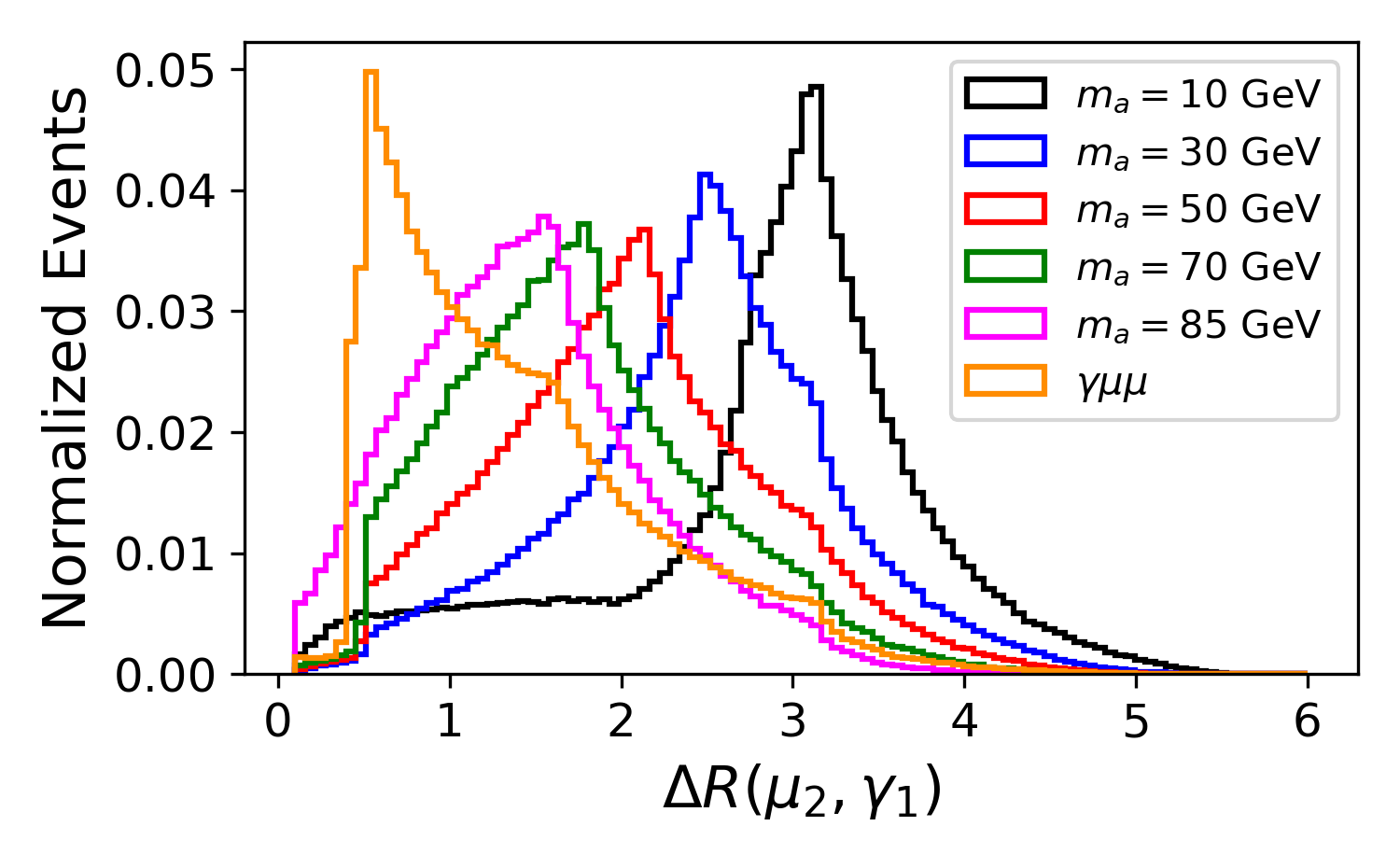}
	\caption{Similar to Fig.\ref{fig:Egammaandma}, but for the angles between the two muons $\Delta R (\mu_1,\mu_2)$ (left panel) and the $\Delta R (\mu_2,\gamma_1)$ (right panel).}
	\label{fig:deltar}
\end{figure}

\begin{figure}[t!]
	\centering
	\includegraphics[width=0.45\linewidth]{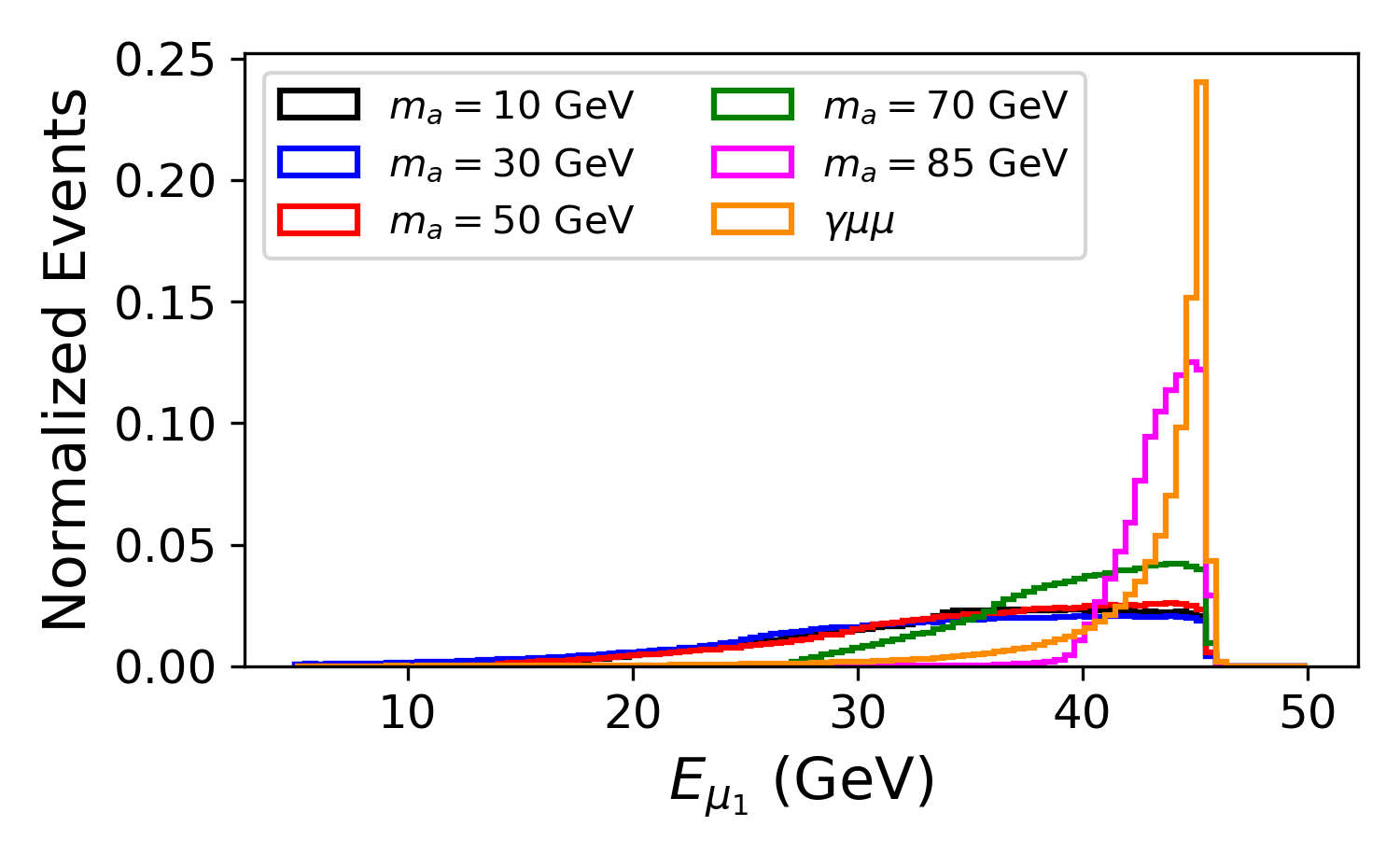}
	\includegraphics[width=0.45\linewidth]{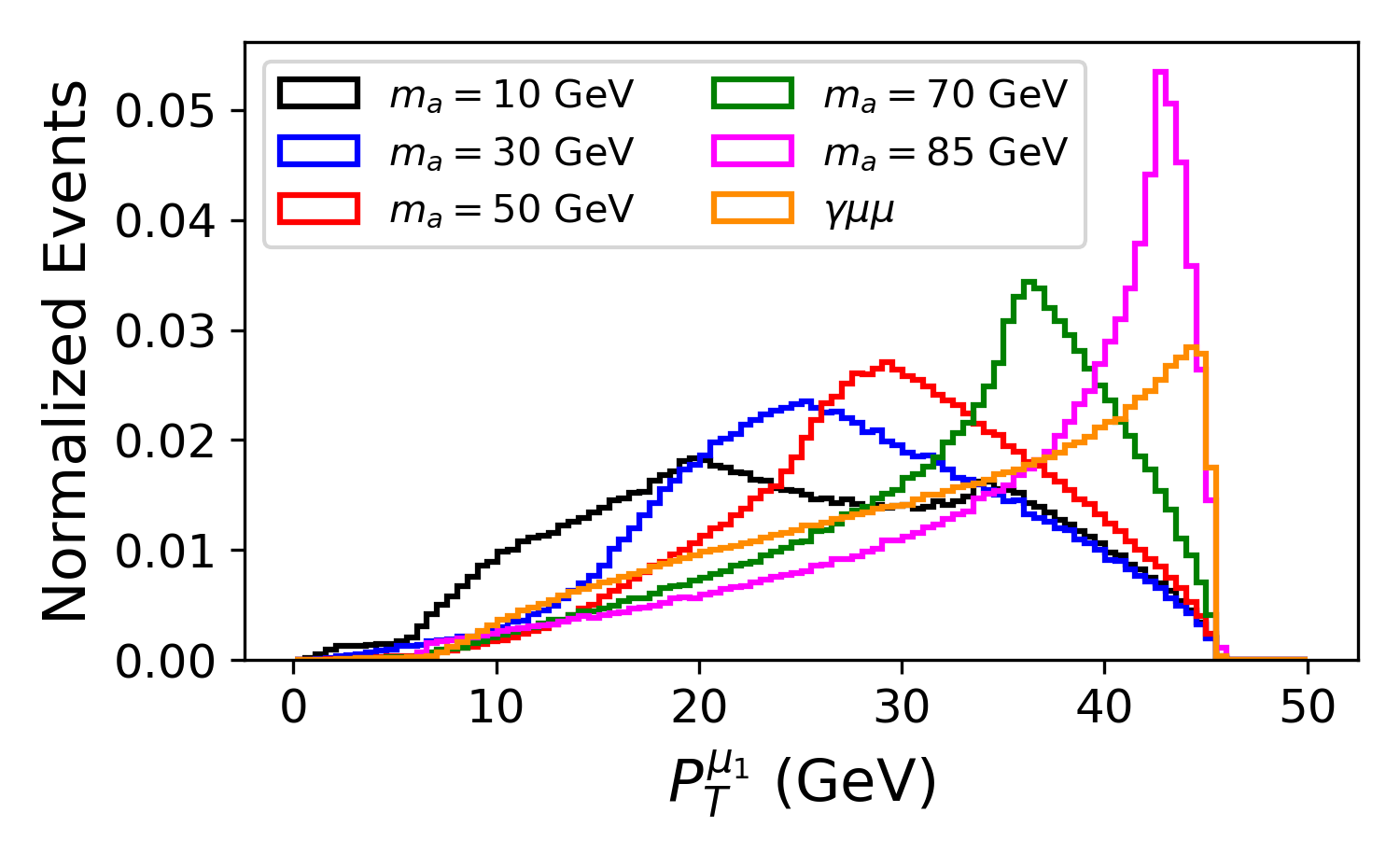}\\
	\includegraphics[width=0.45\linewidth]{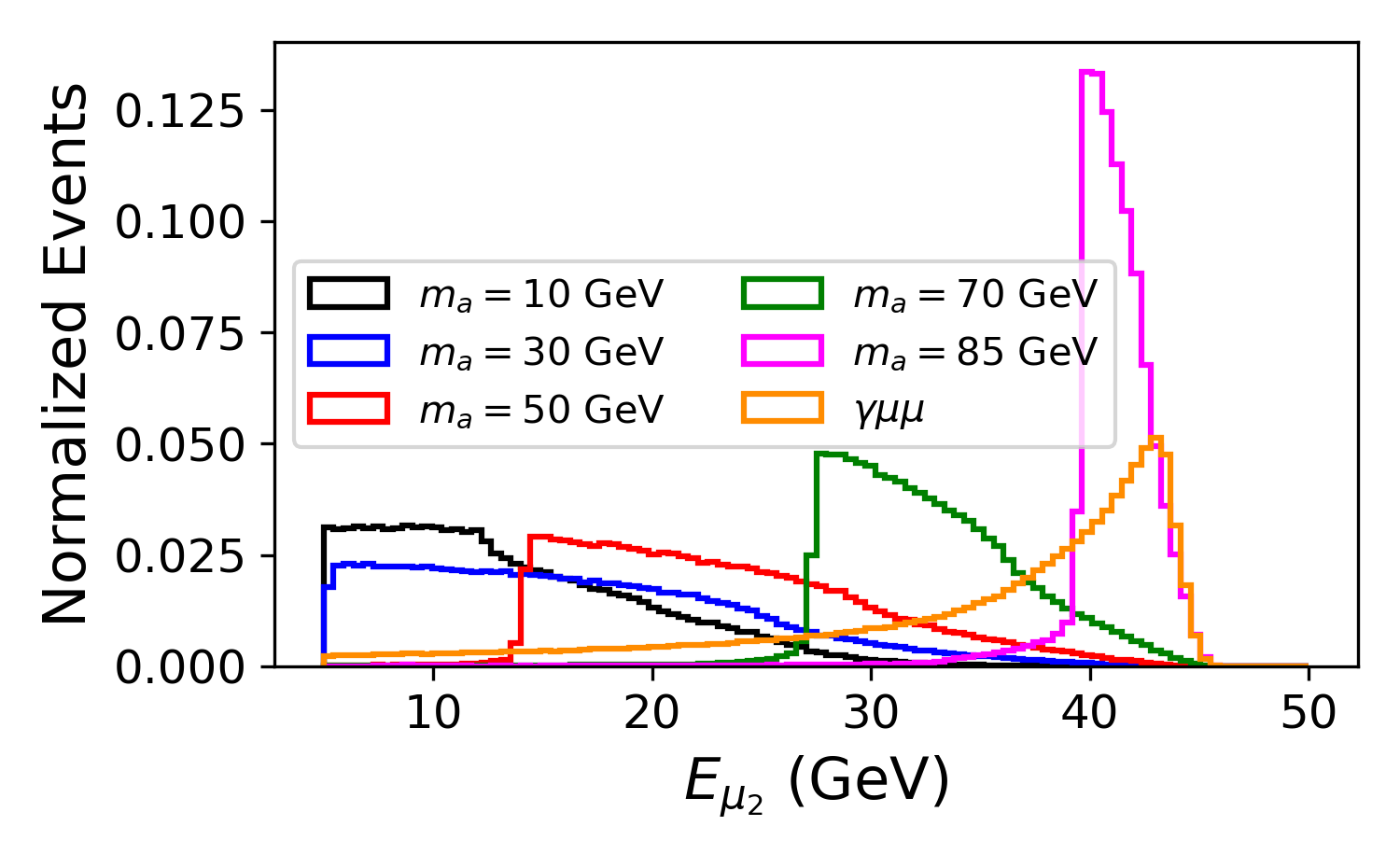}
	\includegraphics[width=0.45\linewidth]{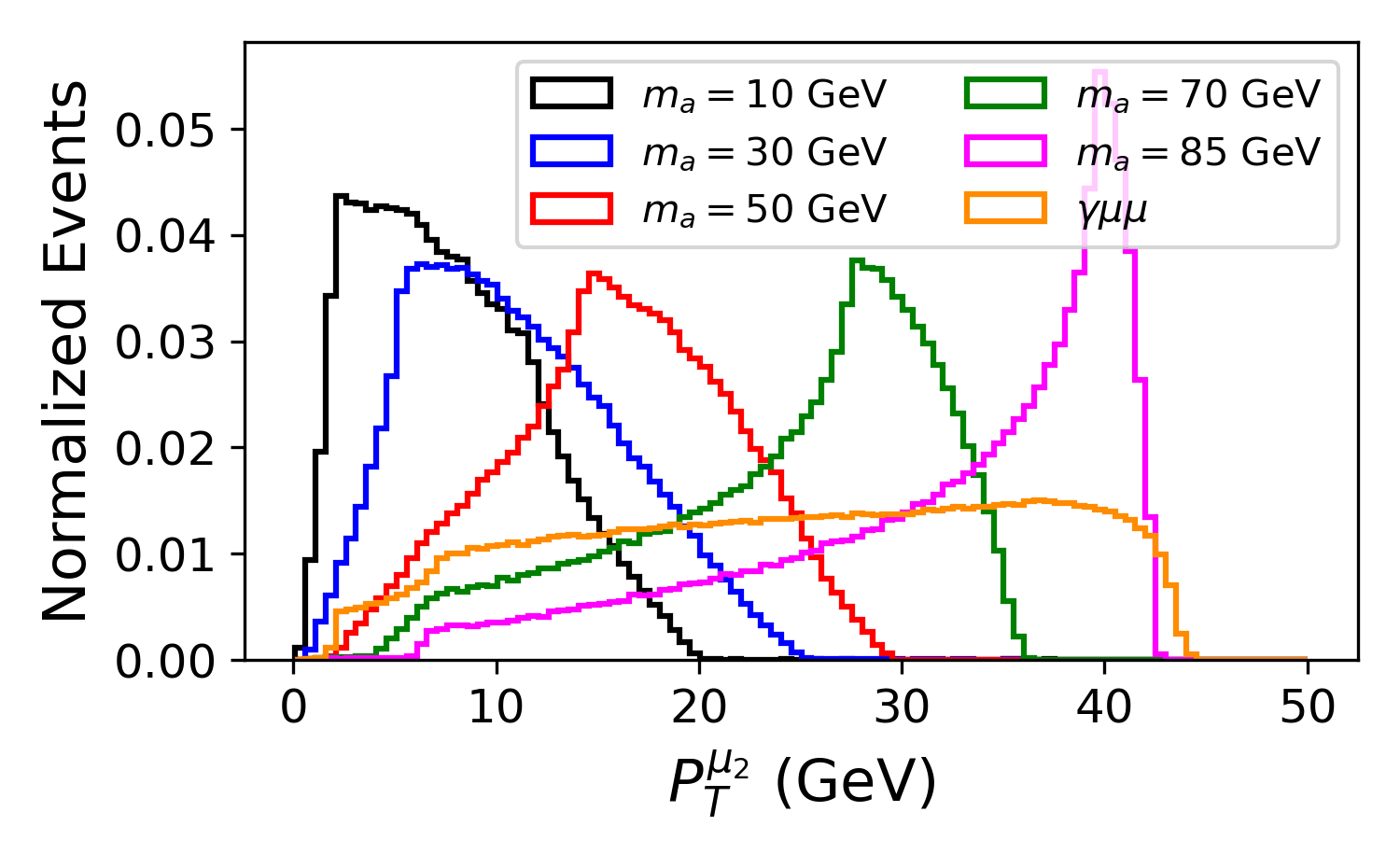}\\
	\includegraphics[width=0.45\linewidth]{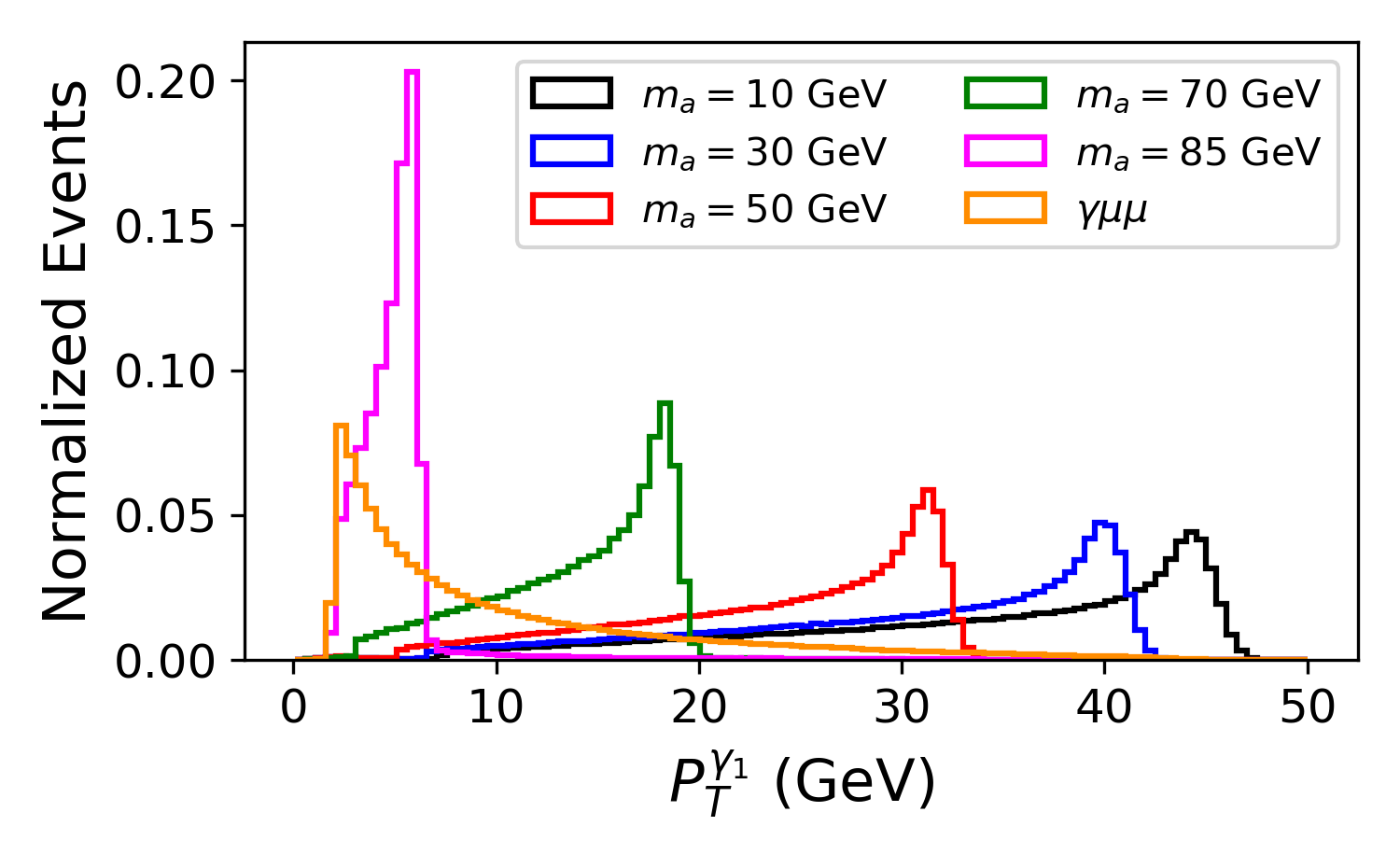}
	\caption{Similar to Fig.\ref{fig:Egammaandma}, this figure shows the energy and transverse momentum distributions of the final-state $\mu_1$ and $\mu_2$, as well as the transverse momentum distribution of the photon. }
	\label{fig:EandPT}
\end{figure}

The final-state muons $\mu_1$ and $\mu_2$ originate from the decay of the ALP and thus exhibit a characteristic angular separation $\Delta R$, which can be approximated by the empirical relation \cite{Reuter:2014iya}:
\begin{eqnarray}
	\Delta R(\mu_{1},\mu_{2}) \propto \frac{2m_{a}}{p_{T}^{a}}\,,
\end{eqnarray}
where $m_a$ and $p_T^a$ denote the ALP mass and transverse momentum, respectively. This relation is illustrated in the left panel of Fig.\ref{fig:deltar}, while the right panel shows the angular separation between $\mu_2$ and the photon. For different ALP mass benchmarks, the signal and background exhibit clearly distinguishable $\Delta R$ distributions, providing strong discriminating power. Additionally, the transverse momentum and energy of the muons, as well as the photon energy, also offer significant separation between signal and background, particularly in the low-mass ALP regime. These distributions are shown in Fig.\ref{fig:EandPT}.

We use XGBoost \cite{Chen:2016btl} as the classifier to distinguish signal events from background. As a state-of-the-art implementation of gradient-boosted decision trees (GBDT), XGBoost offers high performance in separating complex patterns in high-dimensional feature space. The input features for the model are based on the distributions shown in Figs.\ref{fig:Egammaandma}–\ref{fig:EandPT}, made of the kinematic variables of the visible final-state particles.

Several hyperparameters are crucial to the performance of the XGBoost model:
\begin{itemize}
	\item \textbf{Learning rate}: Set to 0.02 to balance stability and convergence speed.
	\item \textbf{Maximum tree depth}: Chosen to be twice the number of input features to ensure sufficient model complexity.
	\item \textbf{Number of trees}: Fixed at 300 to avoid underfitting or overfitting.
	\item \textbf{Regularization}: Kept at the default setting to control model complexity.
\end{itemize}

The classifier performance for the $\mu^+\mu^-$ channel is shown in the left panels of the summary Fig.\ref{fig:ML_performance}. The top two panels illustrate how the classifier can efficiently distinguish between signal and background, with a signal significance increasing towards a high minimum cut on the classifier output. The bottom panel shows the signal and background selection efficiencies as functions of the classifier threshold. For a threshold of $0.97$, background efficiency is reduced to $1.3\%$, while signal efficiency ranges from $31.4\%$ to $96.4\%$, depending on the ALP mass. These results, summarized in Table~\ref{tab:eff}, demonstrate that the classifier significantly improves the signal-to-background ratio without sacrificing signal acceptance.

\begin{table}[h!]
	\centering
	\begin{tabular}{cccccccc}
		\hline
		$m_{a}$[GeV]& Threshold & \textbf{10} & \textbf{30} & \textbf{50} & \textbf{70} & \textbf{85} & BG\\
		\hline
		$\mu^{+}\mu^{-}$& 97\% & 96.4\% & 91.2\% &84.9\% & 69.1\% & 31.4\% & 1.3\%\\
		\hline
	\end{tabular}
	\caption{The table shows the efficiencies of signal and background for muon final state, as well as the predicted values of the machine learning model corresponding to the optimal significance.}
	\label{tab:eff}
\end{table}

\subsection{$\tau^{+}\tau^{-}$ channel}

For the $\tau^+\tau^-$ channel, we consider leptonic $\tau$ decays into $e$ or $\mu$ accompanied by two neutrinos. As a result, the final state contains either two electrons, two muons, or one electron and one muon, along with four neutrinos and a photon. At the detector level, we observe the kinematic properties of the charged leptons and the photon, including their transverse momentum, pseudorapidity, and azimuthal angle.

\begin{figure}
	\centering
	\includegraphics[width=0.45\linewidth]{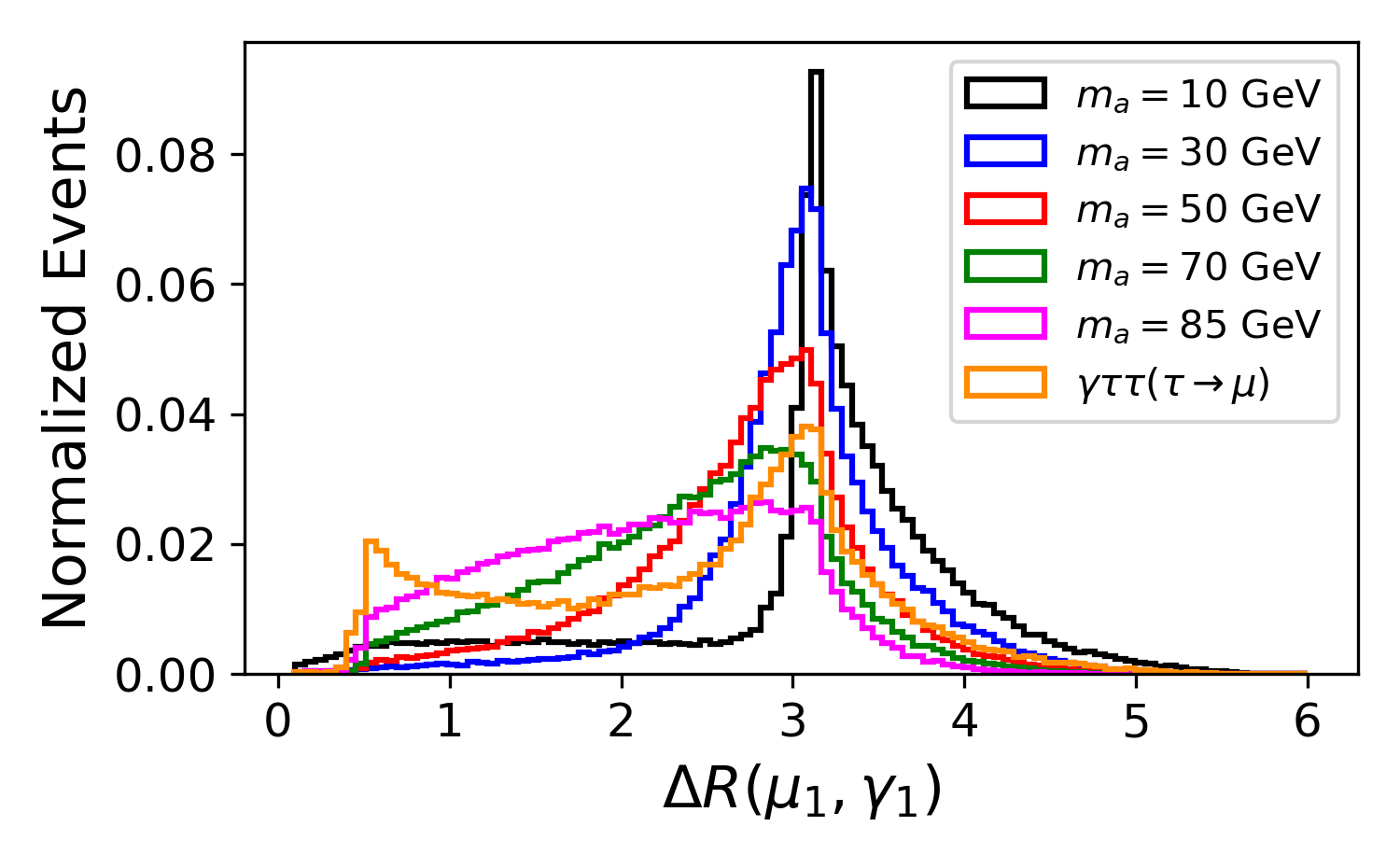}
	\includegraphics[width=0.45\linewidth]{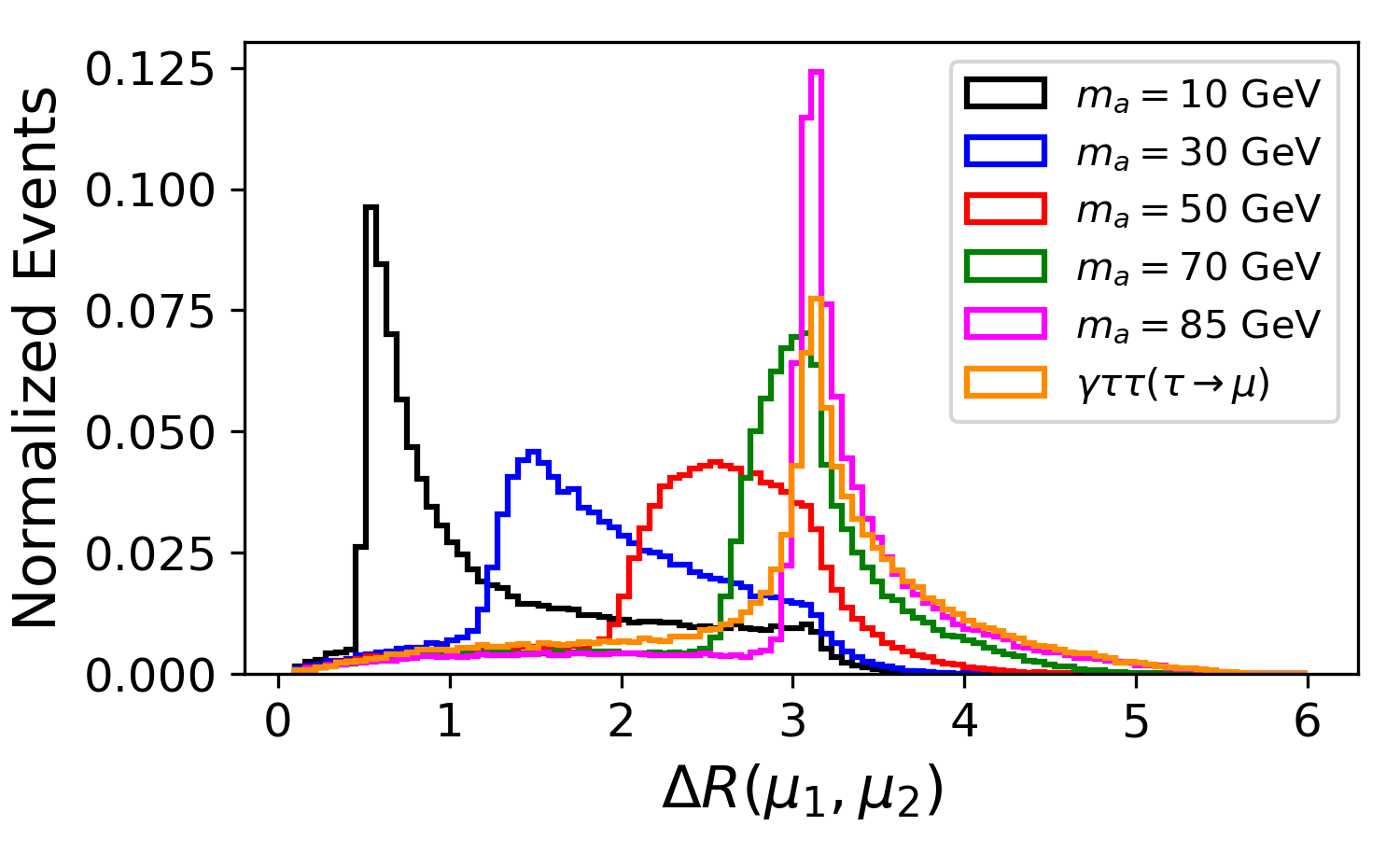}\\
	\includegraphics[width=0.45\linewidth]{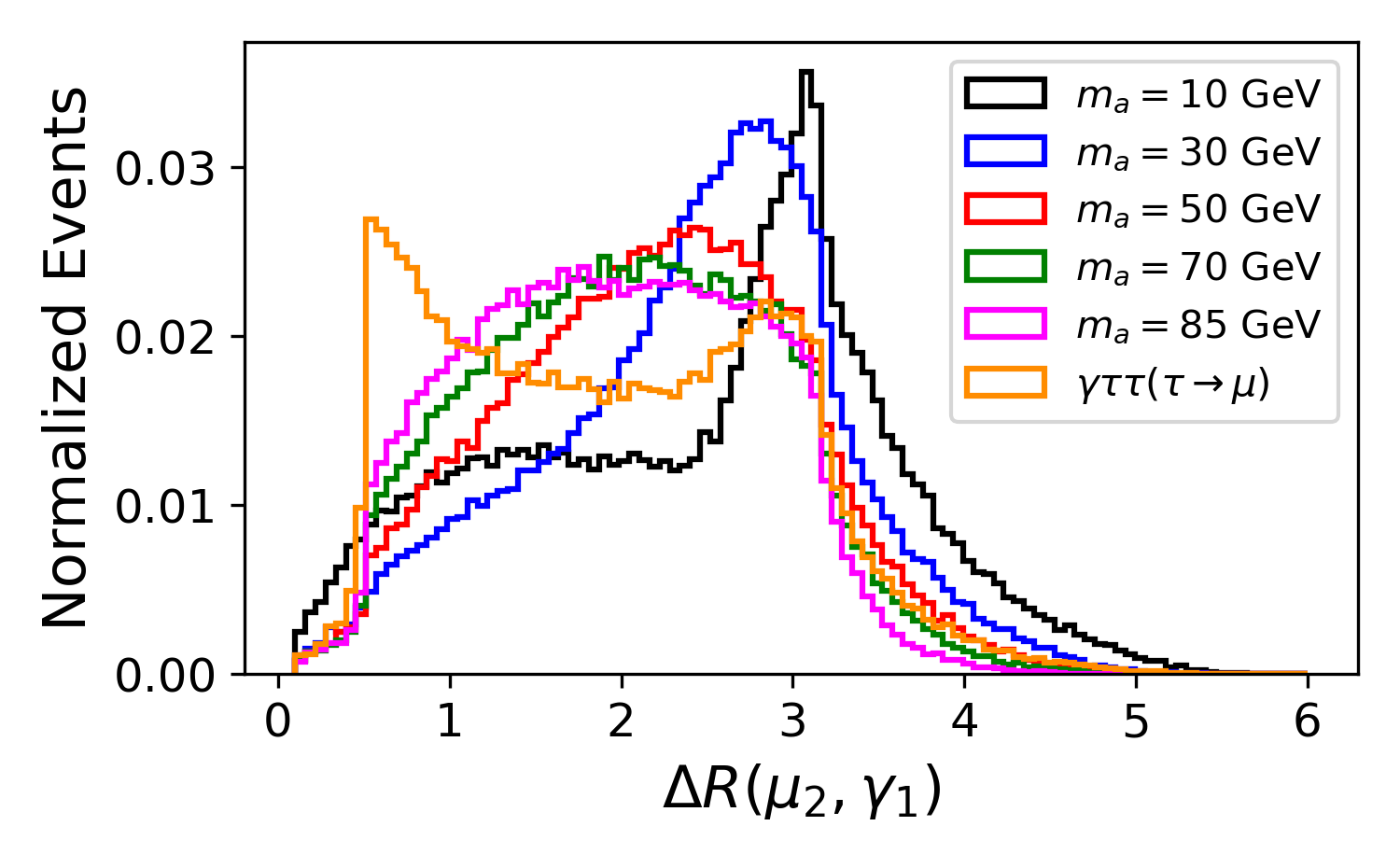}
	\includegraphics[width=0.45\linewidth]{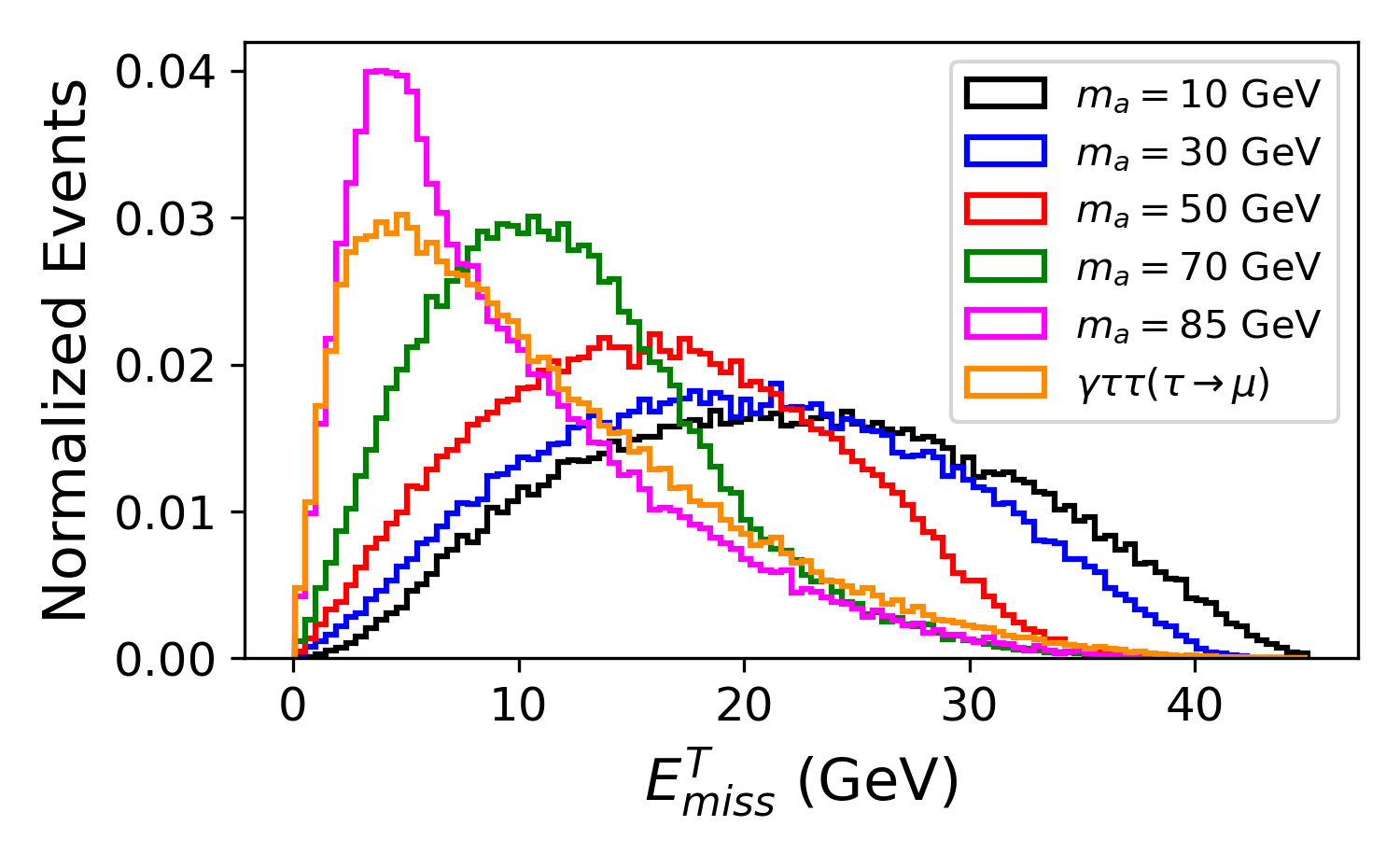}\\
	\includegraphics[width=0.45\linewidth]{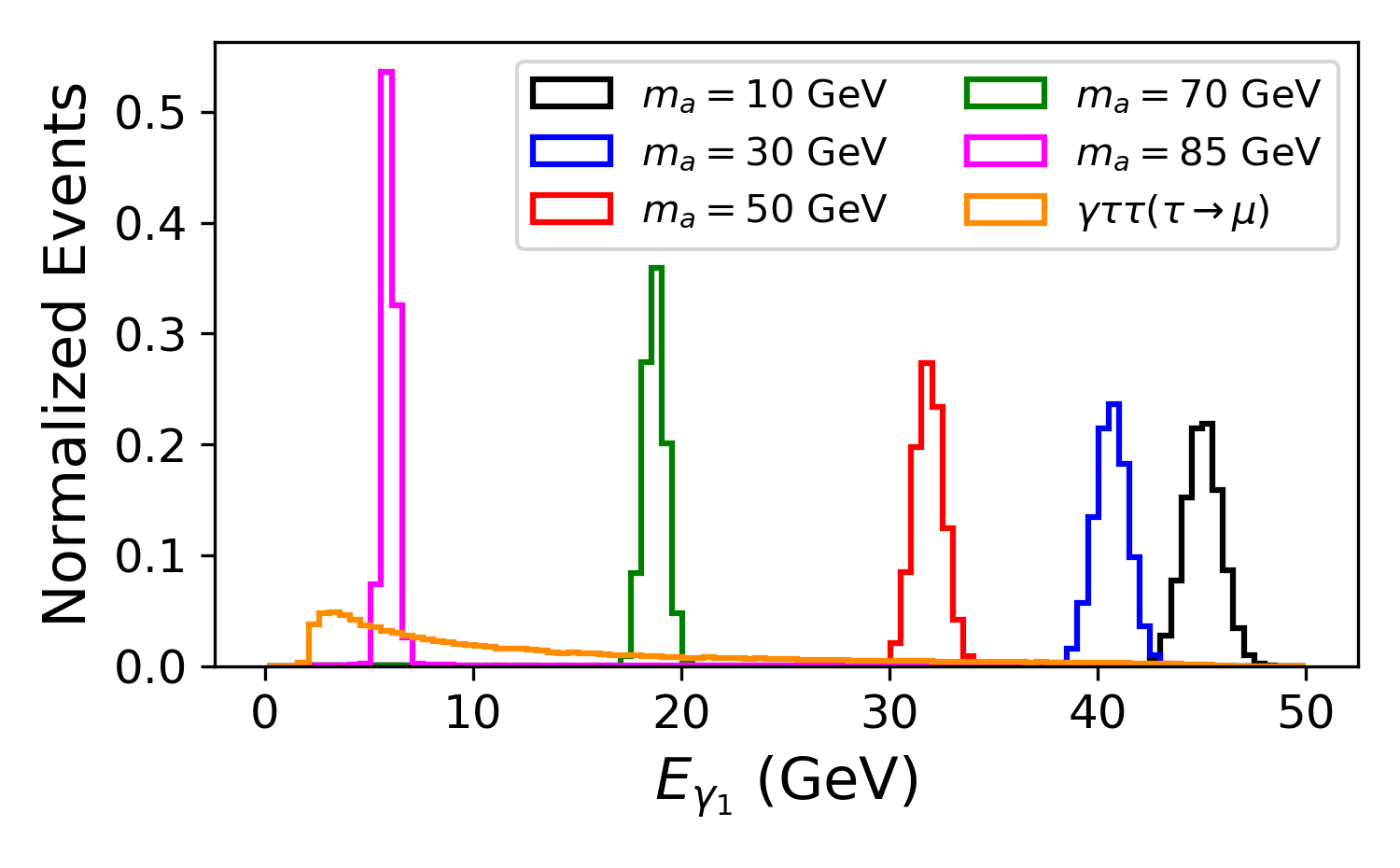}
	\includegraphics[width=0.45\linewidth]{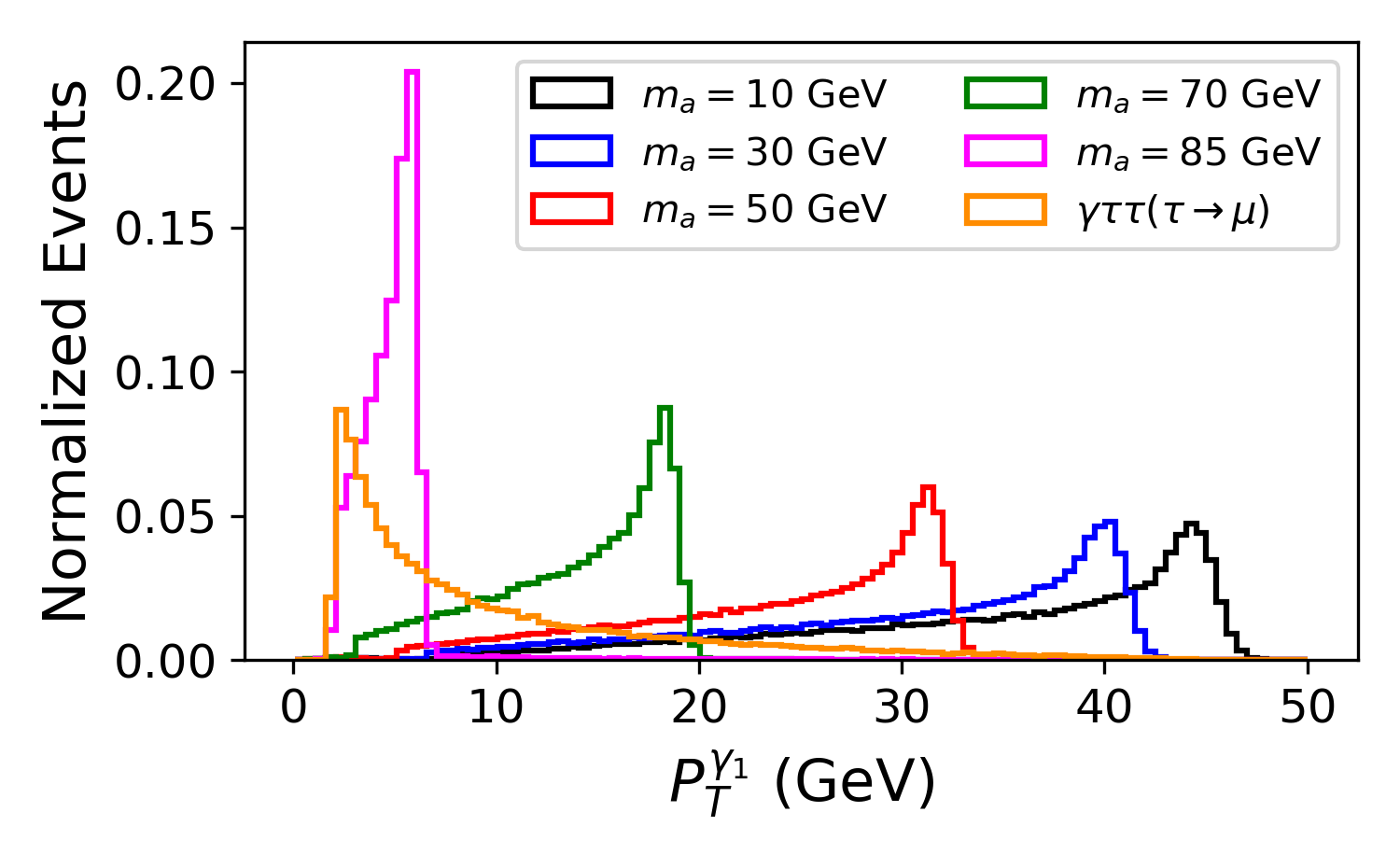}
	\caption{Kinematic distributions for the muon final state, including the photon energy and transverse momentum, transverse missing energy, and angular separations between visible final-state particles. }
	\label{fig:mu}
\end{figure}

The presence of four invisible neutrinos leads to significant transverse missing energy ($E^{T}_{\text{miss}}$), which is a distinguishing feature compared to the $\mu^+\mu^-$ channel. Furthermore, the ALP mass cannot be fully reconstructed in this case, making the photon energy and transverse momentum—determined by the recoil from $Z\to\gamma a$—the most informative observables.

In addition, the angular separations between the final-state leptons ($\Delta R(e_1,e_2)$ or $\Delta R(\mu_1,\mu_2)$ or $\Delta R(\mu,e)$) and between each lepton and the photon ($\Delta R(e_i,\gamma)$ or $\Delta R(\mu_i,\gamma)$) differ significantly between signal and background, particularly in the low-$m_a$ region. These differences enhance the separation power in multivariate analysis.

For all three final states ($ee$, $\mu\mu$, and $e\mu$), we use a common set of six kinematic input variables for the XGBoost classifier. The distributions of these observables are nearly independent on the flavor of the final state leptons, as an example we shown in Fig.\ref{fig:mu} the distributions for the $\mu\mu$ channel. Separate classifiers are trained for each channel, using the same hyperparameter settings as in the $\mu^+\mu^-$ case. The resulting classification performance is nearly independent on the lepton flavors, and we present the $\mu\mu$ case in the central panels of Fig.\ref{fig:ML_performance}, and the corresponding selection efficiencies are summarized in Table~\ref{tab:tau}.

\begin{table}[h!]
	\centering
	\begin{tabular}{cccccccc}
		\hline
		$m_{a}$[GeV]& Threshold & \textbf{10} & \textbf{30} & \textbf{50} & \textbf{70} & \textbf{85} & BG\\
		\hline
		$\tau\to e$& 96.9\% & 92.1\% & 80.7\% & 75.6\% & 59.9\% & 23.4\%  &4\%\\
		$\tau\to\mu$& 96.9\% & 92.3\% &79.7\% & 75.5\% &62.3\%  & 23.9\% &4\%\\
		$\tau\to\mu,\tau\to e$& 96.7\% & 62\% & 76.8\% &73.2\%  & 58.1\% & 16.2\% &3.5\%\\
		\hline
	\end{tabular}
	\caption{The table shows the efficiencies of signal and background for $\tau^{+}\tau^{-}$ final states, as well as the predicted values of the machine learning model corresponding to the optimal significance.}
	\label{tab:tau}
\end{table}

\subsection{Mono-$\gamma$ Signal}

As shown in Fig.\ref{cs_prompt}, ALP particles with masses below 10 GeV predominantly manifest as Mono-$\gamma$ signal at colliders. In such events, the only observable quantities are the kinematic variables of the final-state photon and the transverse missing energy. Accordingly, we consider the $\gamma\nu\bar{\nu}$ process as the dominant background, with a production cross section of $14$~pb. At parton level, we require the photon transverse momentum to exceed $2$~GeV and its pseudorapidity to satisfy $|\eta_{\gamma}| < 2.5$.

\begin{figure}
	\centering
	\includegraphics[width=0.32\linewidth]{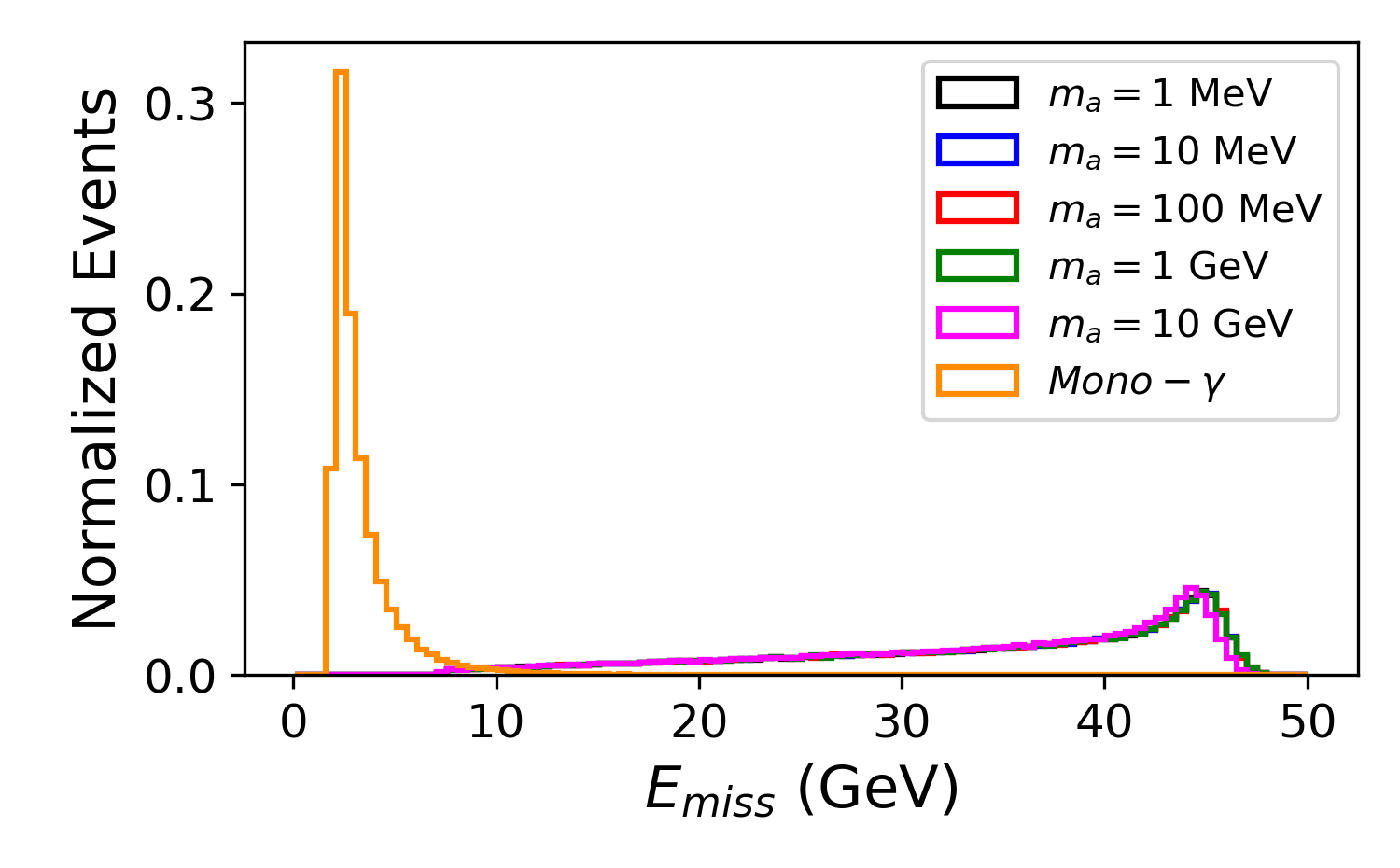}
	\includegraphics[width=0.32\linewidth]{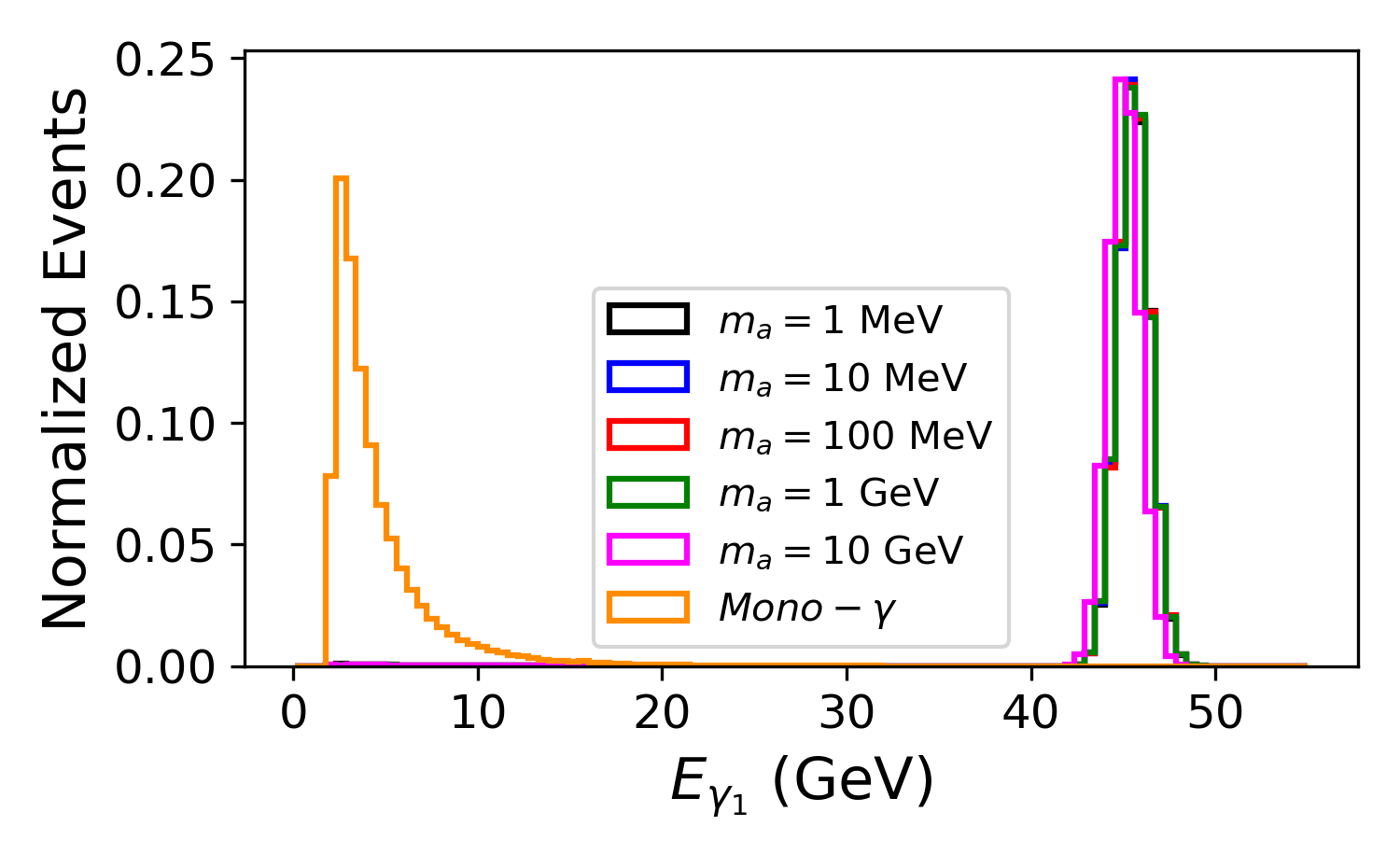}
	\includegraphics[width=0.32\linewidth]{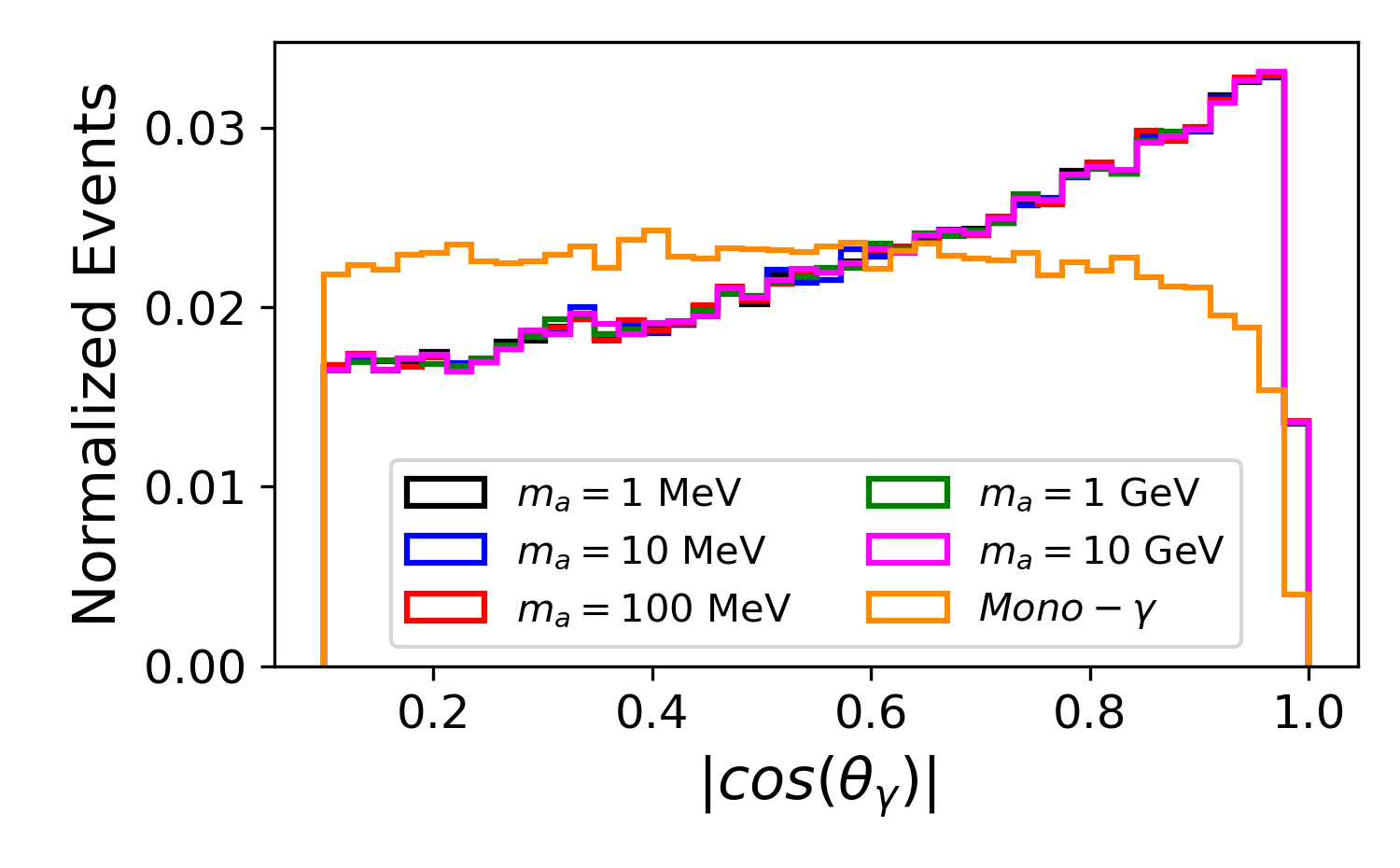}
	\caption{The kinematic distributions of the Mono-$\gamma$ signal, including the photon energy, $|\cos(\theta_{\gamma})|$, and the transverse missing energy. }
	\label{fig:ME}
\end{figure}

\begin{figure}
	\centering
	\includegraphics[width=0.32\linewidth]{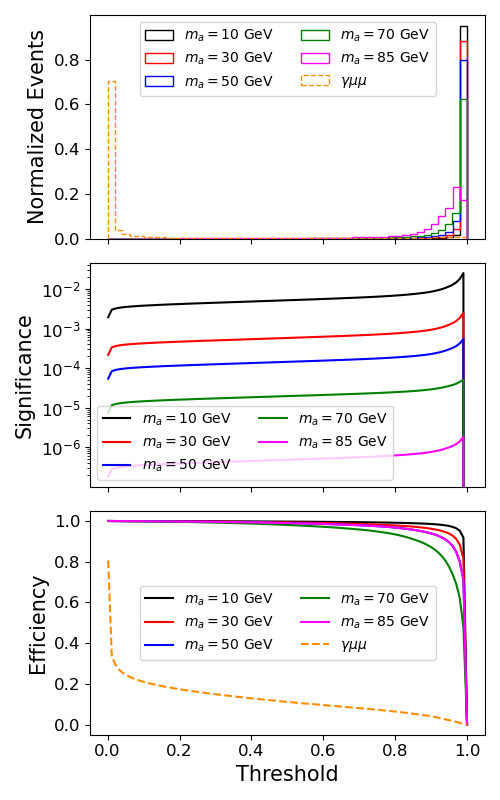}
	\includegraphics[width=0.32\linewidth]{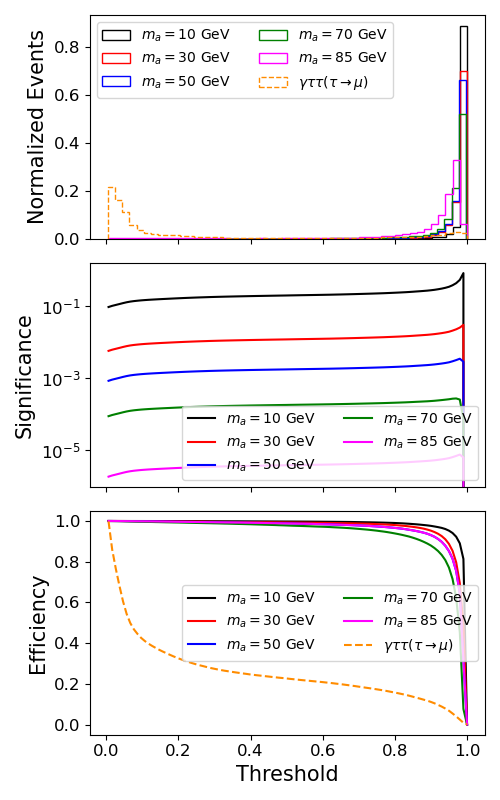} \includegraphics[width=0.32\linewidth]{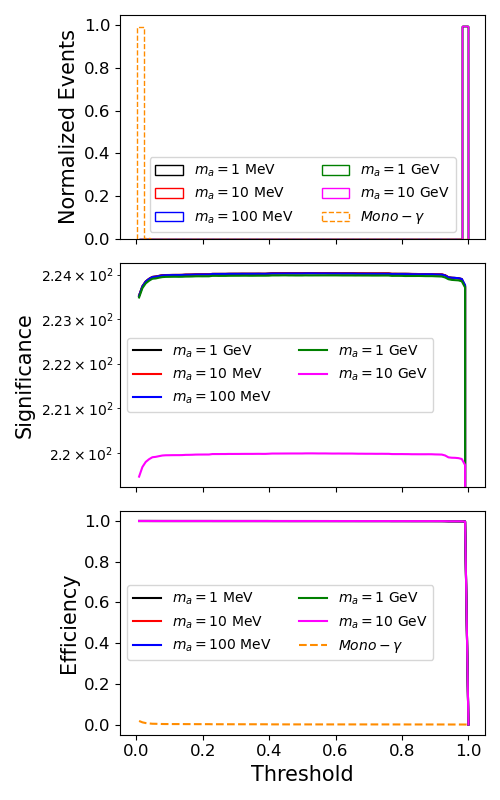}
	\caption{Performance of the XGBoost classifier for the three signals: $a\to\mu^+\mu^-$ (left), $a\to\tau^+\tau^-$ in muons (center), Mono-$\gamma$ (right). The top panels show the predicted distributions for signal (solid line) and background (dashed line) events as a function of the XGBoost output, where the signal peaks near 1 and the background near 0. The middle panels present the signal significance, computed using the formula $Z = S/\sqrt{S + B}$ with 1 ab$^{-1}$, as a function of a cut in the output. The significance reaches its maximum at thresholds close to 1, where the background is highly suppressed while the signal remains relatively intact. The bottom panels display the selection efficiency, defined as the fraction of events (signal or background) passing a given minimum threshold relative to the total. This curve shows how signal and background efficiencies change as the threshold increases. }
	\label{fig:ML_performance}
\end{figure}

To efficiently discriminate signal from background, we use the photon energy, $|\cos(\theta_{\gamma})|$, and transverse missing energy as input features to an XGBoost classifier. The corresponding kinematic distributions are shown in Fig.\ref{fig:ME}. The output of the trained model is displayed in the right panels of Fig.\ref{fig:ML_performance}, where signal and background are well separated with peaks near 0 and 1, respectively, demonstrating excellent discriminative performance. The selection efficiencies for signal and background at a 90\% decision threshold, for various benchmark points, are summarized in Table~\ref{tab:eff_ME}. 

\begin{table}[h!]
	\centering
	\begin{tabular}{cccccccc}
		\hline
		$m_{a}$[GeV]& Threshold & \textbf{10$^{-3}$} & \textbf{10$^{-2}$} & \textbf{10$^{-1}$} & \textbf{1} & \textbf{10} & BG\\
		\hline
		Mono-$\gamma$ & 90\% & 99.9\% & 99.9\% & 99.9\% & 99.9\% & 99.9\%  &0.038\%\\
		\hline
	\end{tabular}
	\caption{The table shows the efficiencies of signal and background for Mono-$\gamma$ signal, as well as the predicted values of the machine learning model corresponding to the optimal significance.}
	\label{tab:eff_ME}
\end{table}

\section{Sensitivities}
\label{result}

The sensitivity to ALPS of FCC-ee and/or CEPC depends on multiple factors, including the choice of model parameters ($C_{WW}$, $C_{BB}$, $C_{GG}$, $c_{ff}$), the ALP mass $m_a$ and coupling scale $\Lambda$, the collider energy, integrated luminosity, and detector performance. In this work, we evaluate the discovery reach for ALPs in the photophobic scenario at $\sqrt{s} = 91.2$~GeV. The statistical significance is defined as:
\begin{eqnarray}
	Z = \frac{S}{\sqrt{S+B}}\,,
\end{eqnarray}
where $S$ and $B$ are the signal and background yields after selection, with efficiencies summarized in Tables~\ref{tab:eff} and \ref{tab:tau} for the lepton channels.

\begin{figure}
	\centering
	\includegraphics[width=0.45\linewidth]{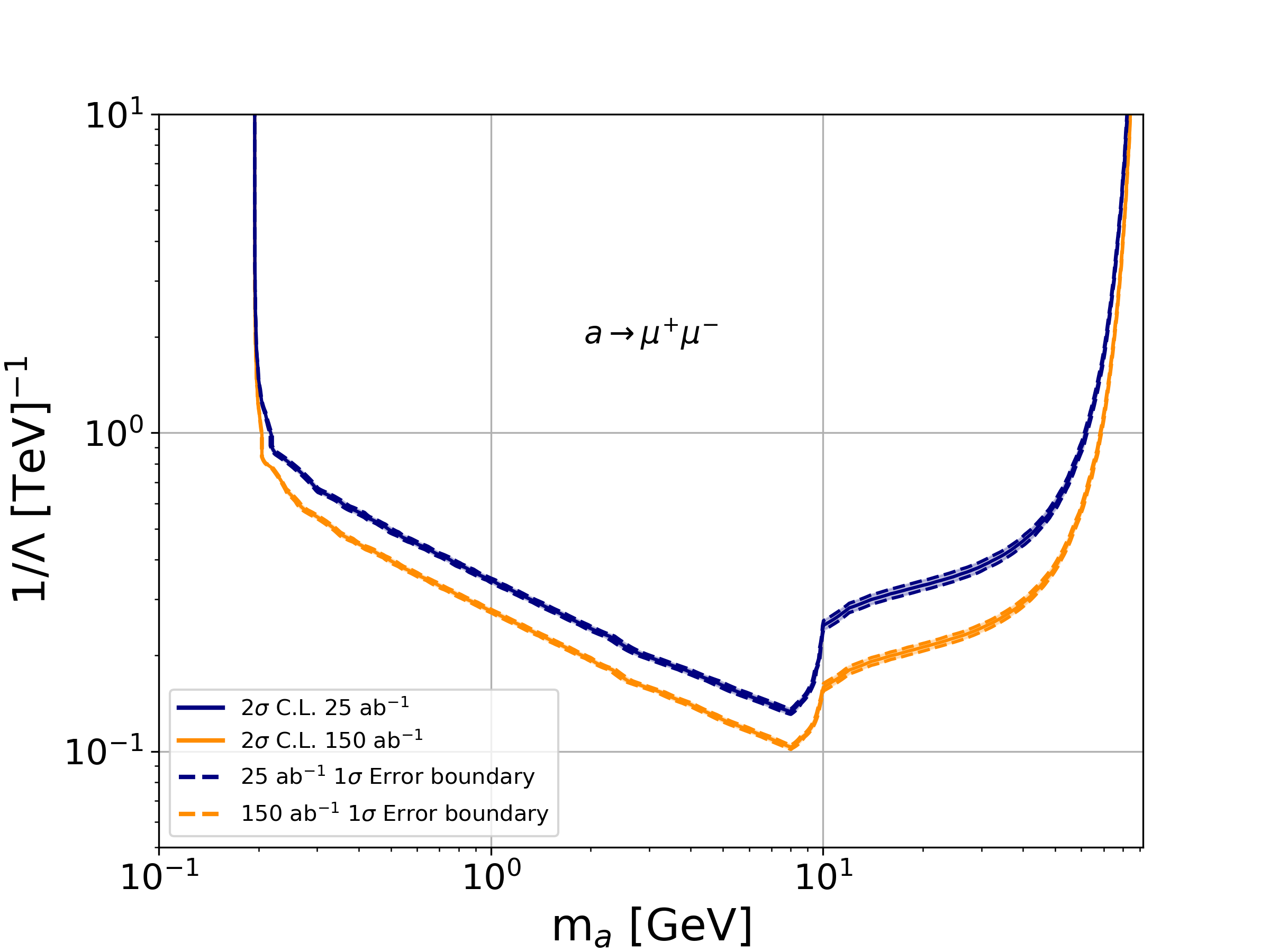}
	\includegraphics[width=0.45\linewidth]{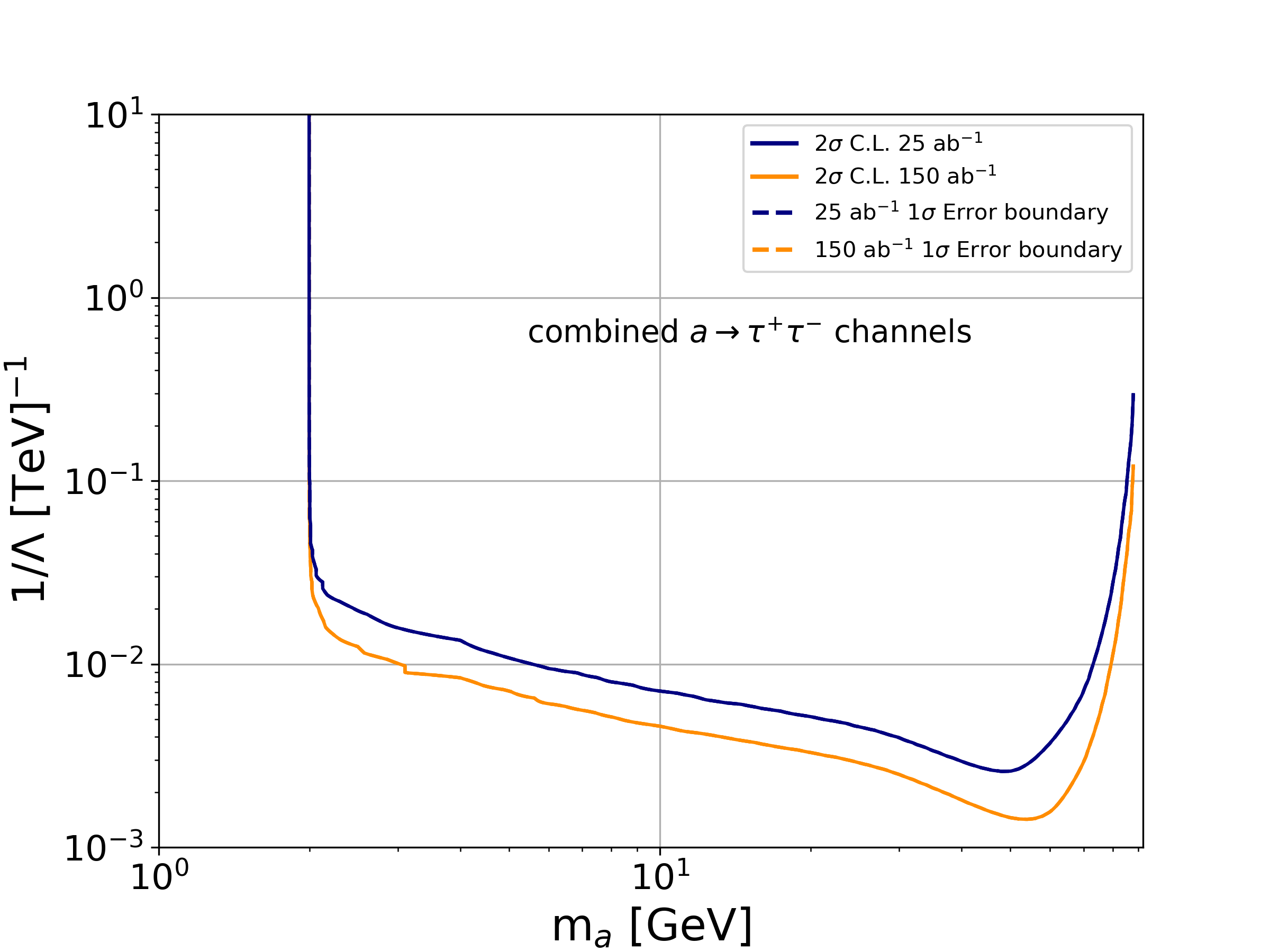}
	\caption{95\% C.L. exclusion limits on $1/\Lambda$ for individual ALP decay channels at $\sqrt{s} = 91.2$~GeV with luminosities of $25$ and $150$~ab$^{-1}$. For the tau channel in the right panel, we combined the three lepton-flavor combinations.}
	\label{fig:amumu}
\end{figure}

We recall that we focused on the photophobic composite ALP, defined by couplings $C_{WW} = - C_{BB} = 1$ and $C_{GG} = c_{ff} = 0$ (where couplings to fermions are generated at loop level).
A parameter scan over the two remaining model parameters, $(m_a, \Lambda)$, is performed using \texttt{EasyScan} \cite{Shang:2023gfy} and sensitivity contours are derived by requiring $Z = 2$, corresponding to a 95\% confidence level exclusion. The exclusion limits for the $\mu^+ \mu^-$ and $\tau^+ \tau^-$ channels are shown in Fig.\ref{fig:amumu} at integrated luminosities of $25$~ab$^{-1}$ (yellow) and $150$~ab$^{-1}$ (darkblue). In both cases, statistical uncertainties are negligible, and the error bands are nearly indistinguishable from the central curves.

From the muon channel in the left panel we see that, due to the impact of the prompt decay probability $P_{\text{prompt}}$, the exclusion sensitivity does not peak at the $2m_\mu$ threshold. As $m_a$ increases, $P_{\text{prompt}}$ grows, enhancing the exclusion reach up to the opening of the $\tau^+\tau^-$ decay channel. Beyond this point, $\text{BR}(a \to \mu^+\mu^-)$ drops sharply, while $P_{\text{prompt}}$ increases, leading to a plateau in sensitivity. Once the $b\bar{b}$ channel opens, the ALP lifetime is dominated by hadronic decays leading to always prompt decays, however the branching ratio falls and the exclusion reach deteriorates. At even higher masses, sensitivity continues to decrease due to the phase space suppression of $\Gamma(Z \to \gamma a) \propto (M_Z^2 - m_a^2)^3$.

The right panel of Fig.\ref{fig:amumu} displays the exclusion limits for the $\tau^+\tau^-$ channel, with the three lepton flavor channels combined statistically by use of the following formula
\cite{Cousins:2007hgg}:
\begin{eqnarray}
	Z_{\mathrm{sum}} = \sum_{i} Z_i, \quad
	\sigma^2\bigl(Z_{\mathrm{sum}}\bigr) = \sum_{i} \sigma^2\bigl(Z_i\bigr), \\
	Z_{\mathrm{comb}} = \frac{Z_{\mathrm{sum}}}{\sqrt{\sigma^2\bigl(Z_{\mathrm{sum}}\bigr)}}\,, \label{eq:Zcomb}
\end{eqnarray}
where the index $i$ runs over the different final states, and Poisson statistics are assumed for both signal and background event counts.
Since the detector performance for electrons and muons is nearly identical, and the branching ratios $\text{BR}(\tau \to e\nu\bar{\nu}) \approx 17.8\%$ and $\text{BR}(\tau \to \mu\nu\bar{\nu}) \approx 17.4\%$ are comparable, the sensitivities from these three cases are nearly identical, c.f. Table~\ref{tab:combine}.
Compared to the $\mu^+\mu^-$ channel, the $\tau^+\tau^-$ channel is less affected by the suppression from $P_{\text{prompt}}$ and benefits from higher branching fractions at moderate $m_a$. As a result, the exclusion reach is significantly improved. In particular, at the $b\bar{b}$ threshold, the drop in the tau branching ratio is compensated by the increase in $P_{\text{prompt}}$ (see Appendix~\ref{appendix}), leading to a smooth exclusion curve.
At higher masses, the sensitivity gradually deteriorates, again due to the reduced phase space for $Z \to \gamma a$.

\begin{figure}
	\centering
        \includegraphics[width=0.7\linewidth]{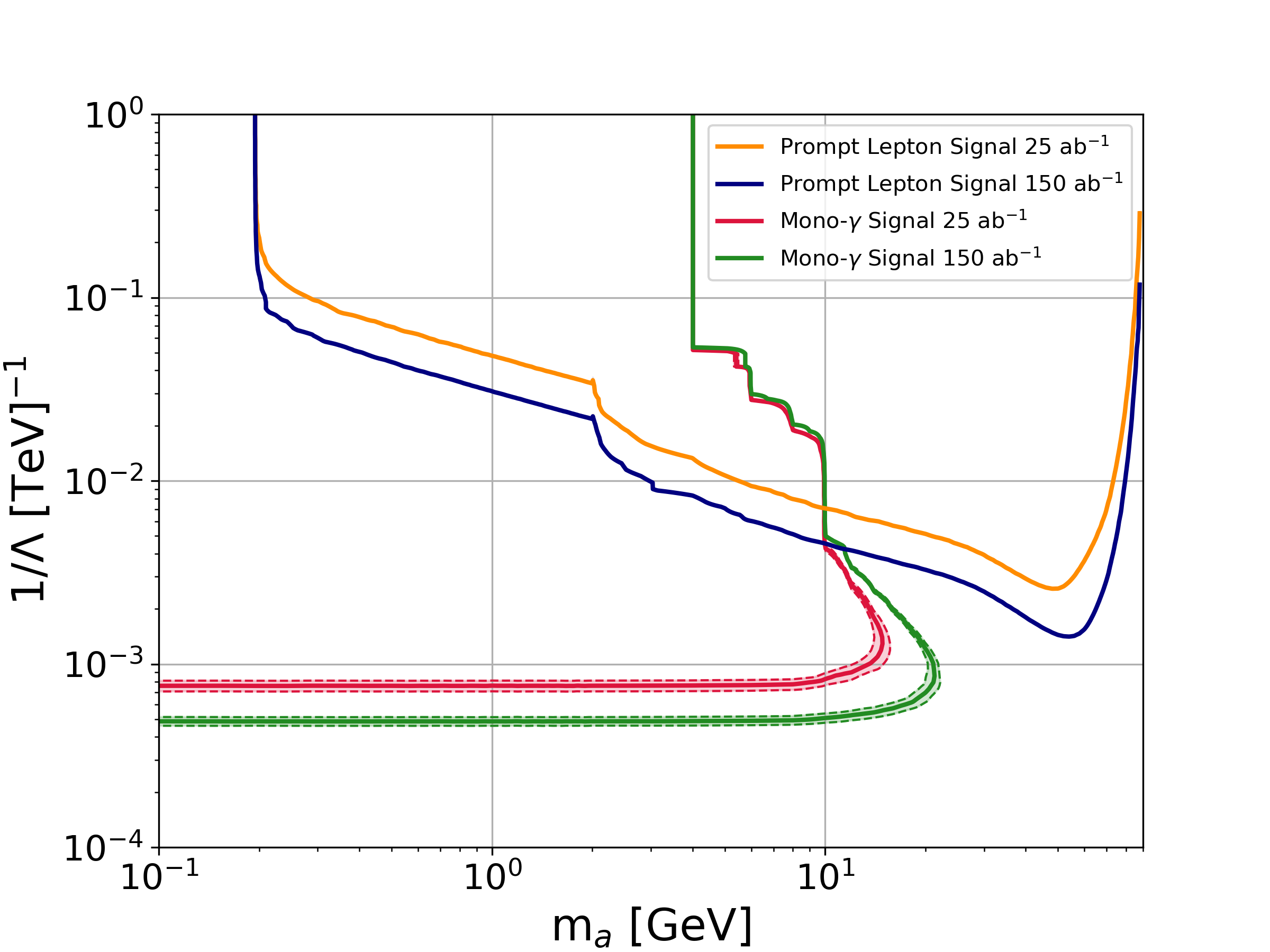}
	\caption{Combined 95\% C.L. exclusion limits on $1/\Lambda$ at $\sqrt{s} = 91.2$~GeV. The dark blue and yellow solid lines represent the exclusion limits after combining all leptonic decay channels at integrated luminosities of  $25$~ab$^{-1}$ and $150$~ab$^{-1}$, respectively. The red and green lines indicate the exclusion limits from the Mono-$\gamma$ signal at the corresponding luminosities. The dashed lines represent the 1$\sigma$ uncertainty bands, which coincide with the central values for the leptonic decay channels.}
	\label{fig:combine}
\end{figure}

In Fig.\ref{fig:combine}, we present the combined exclusion limits from the prompt lepton channels. By combining the four  $\mu^+\mu^-$ and $\tau^+\tau^-$ channels, the sensitivity to the ALP coupling scale $\Lambda$ can be significantly improved. At an integrated luminosity of $150$~ab$^{-1}$, the combined analysis yields an exclusion reach as low as $10^{-3}$~TeV$^{-1}$. A summary of the exclusion limits from each individual channel and the combined result is provided in Table~\ref{tab:combine}, where we also report the results for the Mono-$\gamma$ analysis.

\begin{table}[t!]
	\centering
	\vspace{5pt}
		\begin{tabular}{|c|c|c|c|}
			\hline
			\multirow{2}{*}{$\sqrt{s}$ = 91.2 GeV} & \multicolumn{1}{c|}{25 ab$^{-1}$}&
			\multicolumn{1}{c|}{150 ab$^{-1}$}& \multirow{2}{*}{$m_{a}$  [GeV]}  \\ \cline{2-3}
			&\multicolumn{1}{c|}{$1/\Lambda $ [TeV$^{-1}$] }&
			\multicolumn{1}{c|}{$1/\Lambda $ [TeV$^{-1}$]}& \\ \hline
			$a ~\to~ \mu^{+} \mu^{-}$ & 1.35$\times$10$^{-1}$      &1$\times$10$^{-1}$& [2$m_{\mu}$,91.2]  \\ \hline\hline
			$a ~\to~ \tau^{+} \tau^{-} ~,~ \tau^{+} \to e^{+}\nu_{e}\bar{\nu}_{\tau} ~,~ \tau^{-} \to e^{-}\bar{\nu}_{e}\nu_{\tau}$ & 5.5$\times$10$^{-2}$  & 4$\times$10$^{-2}$ & [2$m_{\tau}$,91.2]  \\ \hline
			$a ~\to~ \tau^{+} \tau^{-} ~,~ \tau^{+} \to e^{+}\nu_{e}\bar{\nu}_{\tau} ~,~ \tau^{-} \to \mu^{-}\bar{\nu}_{\mu}\nu_{\tau}$ & 6$\times$10$^{-2}$& 4$\times$10$^{-2}$ & [2$m_{\tau}$,91.2]  \\ \hline
			$a ~\to~ \tau^{+} \tau^{-} ~,~ \tau^{+} \to \mu^{+}\nu_{\mu}\bar{\nu}_{\tau} ~,~ \tau^{-} \to \mu^{-}\bar{\nu}_{\mu}\nu_{\tau}$& 5.5$\times$10$^{-2}$& 4$\times$10$^{-2}$ & [2$m_{\tau}$,91.2] \\ \hline\hline
			combined $ a  \to \tau^{+} \tau^{-}$& 2.65$\times$10$^{-3}$& 1.45$\times$10$^{-3}$ & [2$m_{\tau}$,91.2] \\ \hline\hline
			combined $ a  \to \mu^{+} \mu^{-} , \tau^{+} \tau^{-}$& 2.6$\times$10$^{-3}$& 1.4$\times$10$^{-3}$ & [2$m_{\mu}$,91.2] \\ \hline\hline
			Mono-$\gamma$& 3.57$\times$10$^{-4}$& 4.8$\times$10$^{-4}$ & [--,21] \\ \hline
		\end{tabular}
\caption{Best sensitivities in the relevant ranges of $m_a$ on $1/\Lambda$ at the 95\% C.L. with $\sqrt{s}=91.2$ GeV. The combined sensitivities are obtained from the statistical formula in Eq.~\ref{eq:Zcomb}.}
\label{tab:combine}
\end{table}

\section{Conclusion}
\label{conclusion}
Axions and axion-like particles have attracted extensive attention well beyond their original role in resolving the strong CP problem. 
This work focuses on the photophobic scenario, specifically motivated by a composite ALP model, exploring the sensitivity of the FCC-ee and CEPC to the process $Z \to \gamma a$ using only leptonic final states.

We focus our attention on prompt decays of the ALP, which apply for decays into $\mu^+\mu^-$ and $\tau^+\tau^-$ pairs, with subsequent tau decays into various lepton flavor combinations. We train separate XGBoost classifiers for each of the four signal processes to distinguish signal from background. The classifiers show excellent performance, as illustrated in Fig.\ref{fig:ML_performance}, offering clear distinction between signal and irreducible backgrounds.
The combined exclusion reach we obtained, shown in Fig.~\ref{fig:combine}, indicates that, at an integrated luminosity of 150~ab$^{-1}$, scales $\Lambda$ between $10$ and $700$~TeV can be probed, for ALP masses between $2m_\mu$ and $75$~GeV. For masses lighter than the $b\bar{b}$ threshold, most decays occur outside the detector, hence offering a complementary Mono-$\gamma$ signature. We apply the same analysis procedure to this channel, with final exclusion reach shown in Fig.~\ref{fig:combine}. Hence, for masses below about $20$~GeV, the scale $\Lambda$ can be probed up to $2\cdot 10^3$~TeV for an integrated luminosity of $150$~ab$^{-1}$ in a parameter space region which is complementary to the prompt lepton channels. A wedge between the regions probed by the two channels, prompt and Mono-$\gamma$, remains open and it could be tested by the search for ALP decays with displaced vertices. However, this feature, characteristic of LLPs, depends crucially on the features of the detector, and we leave its exploration to future work. 

It is important to emphasize that the conclusions of this study apply strictly to the photophobic\cite{Cai:2020bhd,Cai:2019cow} composite scenario, defined by non-vanishing couplings to the electroweak gauge bosons only at the scale $\Lambda$, i.e. $C_{WW} = 1$, $C_{BB} = -1$, $C_{GG} = 0$, and $c_{ff} = 0$. In this model, $\Lambda$ is related to the Higgs compositeness scale $\Lambda \sim 16 \pi^2 f_H$, so that the channels we studied here can be reinterpreted in a reach on $f_H$ between $4.5$ to $13$~TeV, for the prompt leptonic and Mono-$\gamma$ channels, respectively, well beyond the projected reach from Higgs properties and electroweak precision tests.
In other photophobic models, the  different coupling structure would alter the ALP decay modes as well as the behavior of the prompt decay probability $P_{\text{prompt}}$, potentially leading to significantly different reach in the parameter space of the model.

In conclusion, we have shown that prompt lepton decays and Mono-$\gamma$ signatures are extremely effective in probing the parameter space of photophobic ALP models, leaving only a small wedge open, which could be probed by displaced vertex signatures. This study provides an additional strong motivation for a high-luminosity run at the Z-pole for future lepton colliders, such as FCC-ee and CEPC.

\begin{acknowledgments}
This work is supported by the National Natural
Science Foundation of China(NSFC) under Grant No. 12275367, the Fundamental Research Funds for the Central Universities, and the Sun Yat-Sen University Science Foundation. 
\end{acknowledgments}

\appendix

\section{Effect of the prompt decay probability}
\label{appendix} 

\begin{figure}
	\centering
	\includegraphics[width=0.45\linewidth]{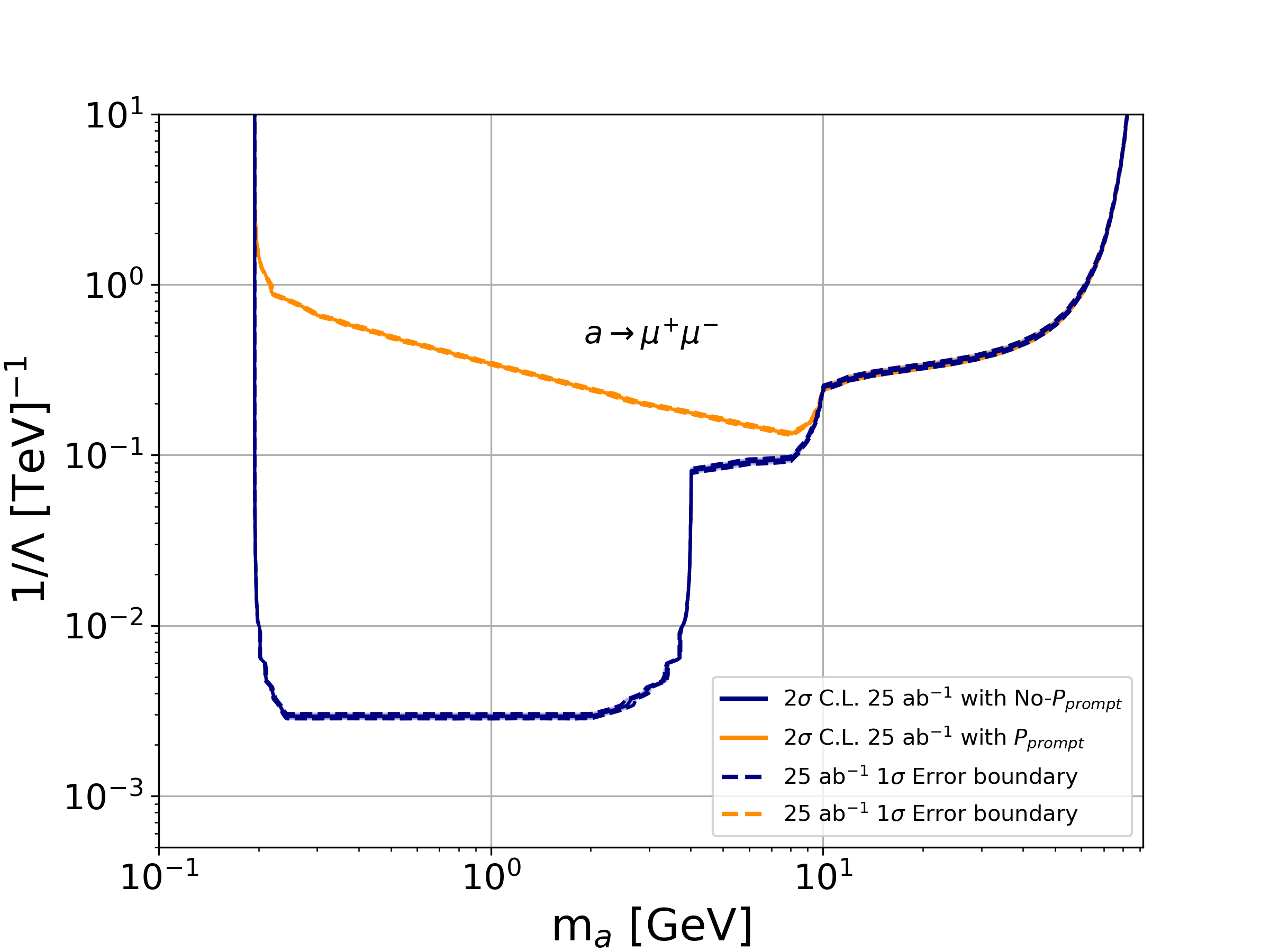}
	\includegraphics[width=0.45\linewidth]{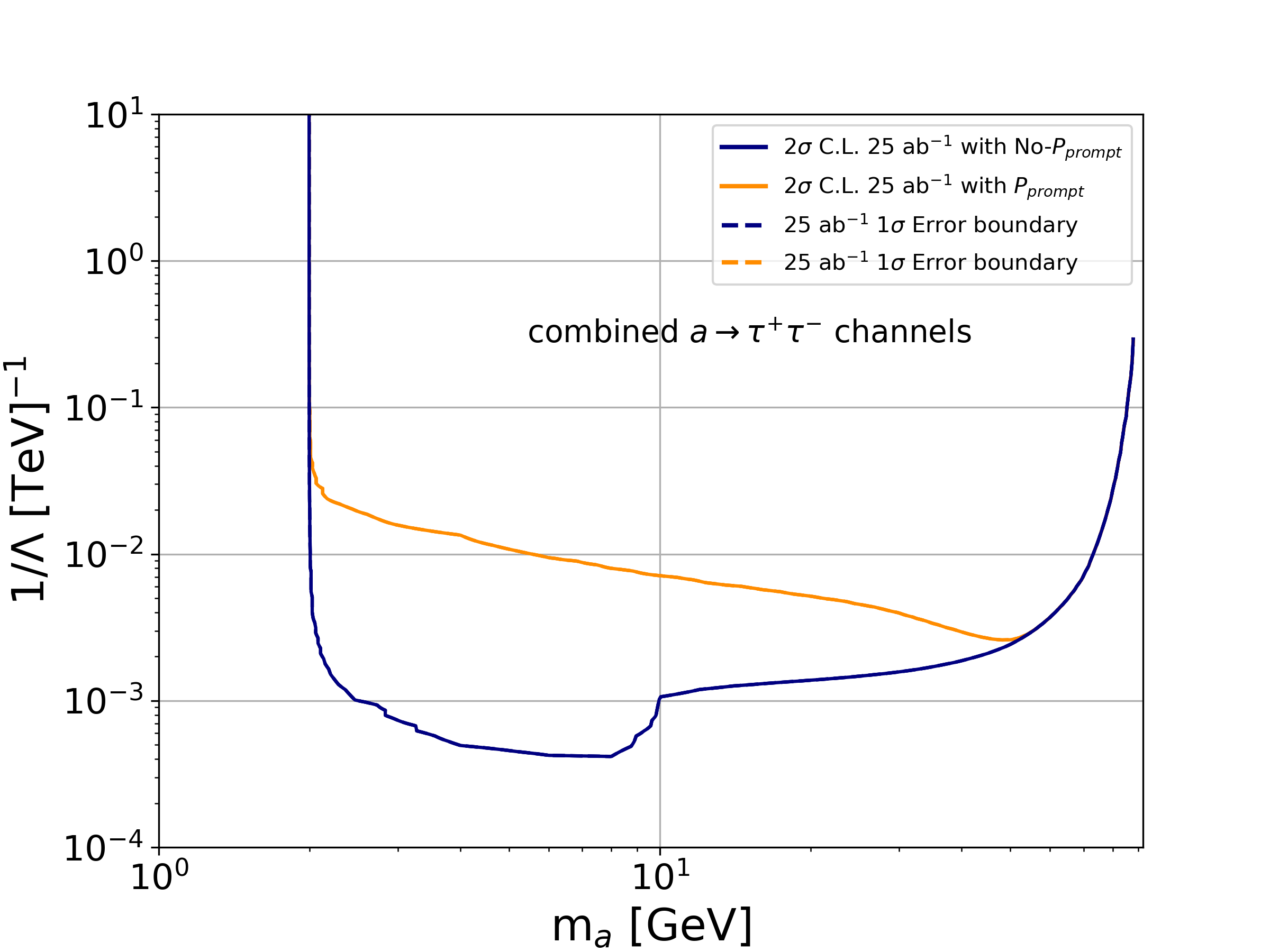}
	\caption{95\% C.L. exclusion limits on $1/\Lambda$ for individual ALP decay channels at $\sqrt{s} = 91.2$~GeV with luminosities of $25$ ~ab$^{-1}$. The blue line assumes 100\% prompt decays, while the orange one considers the actual decay length in the photophobic composite ALP model.}
	\label{prompt_ex}
\end{figure}

In this section, we illustrate the impact of $P_{\mathrm{prompt}}$ on the exclusion limits, as shown in Fig.~\ref{prompt_ex}. The blue solid line represents the exclusion region obtained without considering $P_{\text{prompt}}$ (i.e., assuming $P_{\text{prompt}} = 1$), while the yellow line includes the effect of $P_{\mathrm{prompt}}$.

In the absence of $P_{\text{prompt}}$, the exclusion curve exhibits an upward shift when the ALP becomes kinematically allowed to decay into new channels (i.e., $m_a = 2m_\tau$ or $2m_b$). This is because the branching ratio of the original decay channel sharply decreases once a new channel becomes kinematically allowed. Meanwhile, the opening of new decay modes also shortens the ALP lifetime, leading to a sharp increase in $P_{\mathrm{prompt}}$. As a result, after accounting for both the reduction in the branching ratio and the increase in $P_{\mathrm{prompt}}$, the exclusion limits follow the trend indicated by the yellow curve.

\bibliographystyle{utphys}
\bibliography{ref}

\end{document}